\documentclass[conf,twocolumn]{IEEEtran}
\usepackage{cite}
\usepackage[cmex10]{amsmath}
\usepackage{amsthm}
\usepackage{amssymb}
\usepackage{dsfont}
\usepackage{balance}
\usepackage[final]{graphicx}
\usepackage{color}
\usepackage{epsfig}
\usepackage{epstopdf}
\usepackage{tikz}
\usetikzlibrary{positioning}
\usepackage[hidelinks]{hyperref}
\usepackage{subcaption}
\usepackage[font=footnotesize]{caption}

\newtheorem{theorem}{Theorem}

\newcommand{\bs}[1]{\boldsymbol{#1}}
\newcommand{\mc}[1]{\mathcal{#1}}

\newcommand{\mb}[1]{\mathbf{#1}}
\newcommand{\mr}[1]{\mathrm{#1}}
\newcommand{\ms}[1]{\mathsf{#1}}
\newcommand{\tr}{\mathrm{Tr}}
\newcommand{\E}{\mathbb{E}}
\newcommand{\lr}[1]{\langle #1 \rangle}
\newcommand{\blr}[1]{\big\langle #1 \big\rangle}

\newcommand{\Tp}{T_{\mr{p}}}
\newcommand{\Td}{T_{\mr{d}}}
\newcommand{\Pp}{P_{\mr{p}}}
\newcommand{\xpt}{\mb{x}_{\mr{p},t}}
\newcommand{\xdt}{\mb{x}_{\mr{d},t}}
\newcommand{\xpit}{x_{\mr{p},it}}
\newcommand{\xpjt}{x_{\mr{p},jt}}
\newcommand{\xdit}{x_{\mr{d},it}}

\newcommand{\ypt}{\mb{y}_{\mr{p},t}}
\newcommand{\ydt}{\mb{y}_{\mr{d},t}}
\newcommand{\ypit}{y_{\mr{p},it}}
\newcommand{\ydmt}{y_{\mr{d},mt}}

\newcommand{\ypmt}{y_{\mr{p},mt}}

\newcommand{\rpt}{\mb{r}_{\mr{p},t}}
\newcommand{\rdt}{\mb{r}_{\mr{d},t}}
\newcommand{\rpit}{r_{\mr{p},it}}

\newcommand{\rpmt}{r_{\mr{p},mt}}

\newcommand{\rdmt}{r_{\mr{d},mt}}

\newcommand{\Xp}{\mb{X}_{\mr{p}}}
\newcommand{\Xd}{\mb{X}_{\mr{d}}}
\newcommand{\Yp}{\mb{Y}_{\mr{p}}}
\newcommand{\Yd}{\mb{Y}_{\mr{d}}}

\newcommand{\Rp}{\mb{R}_{\mr{p}}}
\newcommand{\Rd}{\mb{R}_{\mr{d}}}

\newcommand{\gp}{\gamma_{\mr{p}}}
\newcommand{\gdt}{\gamma_{\mr{d},t}}
\newcommand{\low}{\mr{low}}
\newcommand{\up}{\mr{up}}

\DeclareMathOperator*{\argmax}{arg\;max}

\usepackage{mathtools,algorithm}
\usepackage[noend]{algpseudocode}
\def\NoNumber#1{{\def\alglinenumber##1{}\State #1}\addtocounter{ALG@line}{-1}}
\definecolor{LayerColor}{RGB}{230, 230, 230}
\definecolor{InOutColor}{RGB}{240, 243, 255}
\definecolor{cellColor}{RGB}{230, 230, 230}

\begin{document}
	
\title{Variational Bayes for Joint Channel Estimation and Data Detection in Few-Bit Massive MIMO Systems}


\author{Ly~V.~Nguyen,  A.~Lee~Swindlehurst, and Duy~H.~N.~Nguyen
		\thanks{Ly V. Nguyen  and A. Lee Swindlehurst are with the Center for Pervasive Communications and Computing, Henry Samueli School of Engineering, University of California, Irvine, CA, USA 92697 (e-mail: vanln1@uci.edu, swindle@uci.edu).}
		\thanks{Duy H. N. Nguyen is with the Department of Electrical and Computer Engineering, San Diego State University, San Diego, CA, USA 92182 (e-mail: duy.nguyen@sdsu.edu).}}

\maketitle
	
\begin{abstract}
	Massive multiple-input multiple-output (MIMO) communications using low-resolution analog-to-digital converters (ADCs) is a promising technology for providing high spectral and energy efficiency with affordable hardware cost and power consumption. However, the use of low-resolution ADCs requires special signal processing methods for channel estimation and data detection since the resulting system is severely non-linear. This paper proposes joint channel estimation and data detection methods for massive MIMO systems with low-resolution ADCs based on the variational Bayes (VB) inference framework. We first derive matched-filter quantized VB (MF-QVB) and linear minimum mean-squared error quantized VB (LMMSE-QVB) detection methods assuming the channel state information (CSI) is available. Then we extend these methods to the joint channel estimation and data detection (JED) problem and propose two methods we refer to as MF-QVB-JED and LMMSE-QVB-JED. Unlike conventional VB-based detection methods that assume knowledge of the second-order statistics of the additive noise, we propose to float the noise variance/covariance matrix as an unknown random variable that is used to account for both the noise and the residual inter-user interference. We also present practical aspects of the QVB framework to improve its implementation stability. Finally, we show via numerical results that the proposed VB-based methods provide robust performance and also significantly outperform existing methods.
\end{abstract}
	
\begin{IEEEkeywords}
	Approximate message passing, Bayesian inference, detection, estimation, massive MIMO, soft interference cancellation, variational Bayesian.
\end{IEEEkeywords}
	
\section{Introduction}
Beyond-5G wireless systems will require exploitation of the large bandwidths available at THz frequencies (0.3–3 THz)~\cite{Akyildiz2014Terahertz,Akyildiz2014TeraNets,Rajatheva2020white}. An inherent challenge in operating in these bands is the strong radio frequency (RF) path loss, and while this can be effectively addressed by exploiting the beamforming gain available from large antenna arrays, scaling up existing RF technologies to very large arrays becomes complex, expensive, and demands high power consumption. Therefore, implementing massive antenna arrays for THz communications will require radical simplifications in the RF architecture. Hybrid analog-digital arrays reduce the number of RF chains with respect to (w.r.t.) the number of antenna elements~\cite{Heath2016Overview}, but this approach yields poor spatial multiplexing and does not scale well at higher frequencies and wider bandwidths due to the need for complex analog circuitry and resource-consuming beam management schemes~\cite{Saxena2017Analysis}. 

An alternative solution is retaining the RF chains for each antenna, but reducing complexity and energy consumption through the use of low-resolution analog-to-digital converters (ADCs). It has been shown that fully digital arrays with lower-resolution data converters (even down to 1 bit) can significantly outperform hybrid analog-digital architectures in terms of beamforming flexibility and spectral/energy efficiency~\cite{Kilian2018Comparison}. This is because the use of low-resolution ADCs maintains the high spatial multiplexing gains of massive arrays, and they more easily scale to higher frequencies and bandwidths with significantly reduced hardware cost and power consumption. However, the use of low-resolution quantization requires special signal processing methods for channel estimation and data detection since the resulting system is severely non-linear, and the received signals are significantly distorted.

There has been a plethora of channel estimation and data detection studies for massive MIMO systems with low-resolution ADCs. For example, one-bit ML and near-ML methods were proposed in~\cite{choi2016near}. The Bussgang decomposition was used to derive different linear channel estimators in~\cite{li2017channel,Kolomvakis2020Quantized}  and linear data detectors in~\cite{ Lan2018Linearized,Ly2021Linear,Kolomvakis2020Quantized}. While the ML and near-ML methods are either too complicated for practical implementation or non-robust at high signal-to-noise ratios (SNRs), the linear Bussgang-based receivers have lower complexity and are more robust, but they have limited performance. Several other detection approaches have been proposed in~\cite{jeon2019robust,Song2019CRC-Aided,Cho2019OneBitSCSO,Shao2018Iterative} but they require the use of either a cyclic redundancy check (CRC) or an error correcting code (ECC). The authors in~\cite{wen2016bayes} developed a bilinear generalized approximate message passing (BiGAMP) algorithm \cite{BiGAMP-P1-2014} to solve the joint channel estimation and data detection (JED) problem for few-bit MIMO systems.

Recently, machine learning for low-resolution MIMO channel estimation and data detection has gained interest and there has also been numerous results reported in the literature. In particular, the work in~\cite{Ly2021SVM} shows how support vector machine (SVM) models can be applied to one-bit massive MIMO channel estimation and data detection. The authors of~\cite{Ly2021Linear} exploit a deep neural network (DNN) framework to develop a special model-driven detection approach that outperforms the SVM-based methods in~\cite{Ly2021SVM}. Deep learning-based joint pilot signal and channel estimator designs were proposed in~\cite{DuyNguyen2020Neural} and~\cite{Ly2022TWC}. While a conventional DNN structure was used in~\cite{ DuyNguyen2020Neural }, the work in~\cite{Ly2022TWC} employed a model-driven network similar to~\cite{Ly2021Linear}. The work in~\cite{Khobahi2021LoRD-Net} proposed another DNN-based detector but its computational complexity is high since the detection network must be retrained for each new channel realization. Several learning-based blind detection methods were proposed in~\cite{Jeon2018supervised,Ly2020Supervised, Kim2020Machine } but they are restricted to small-scale systems. In~\cite{Xiang2022Bayesian}, Bayesian inference was used to develop a JED method for quantized single-antenna systems with orthogonal frequency division multiplexing (OFDM) and time-frequency doubly selective (DS) channels where the sparsity of the DS channels was exploited. Another JED method was proposed in~\cite{Thoota2021Variational} based on the variational Bayesian (VB) inference framework, and it was shown to outperforms the BiGAMP-based method in~\cite{wen2016bayes} for soft symbol decoding. In a recent work~\cite{Duy-TSP-22}, VB inference was also shown to be very efficient in MIMO data detection with infinite-resolution (perfect) ADCs.

In this paper, we develop a VB framework for channel estimation and data detection for massive MIMO systems with low-resolution ADCs. While conventional machine learning models such as SVM and DNN only provide a point estimate of the signal of interest, e.g., the channel or the data symbols, the VB approach can provide the posterior distribution of the estimate, which is important in subsequent signal processing steps such as channel decoding. Another advantage of VB is that it does not require a training process like DNNs which often suffer from performance degradation due to mismatch between the actual model and that used during training. Unlike our previous work in~\cite{Ly2022Asilomar} which only considers the data detection problem and assumes perfect channel state information (CSI), we study both channel estimation and data detection in this paper and make the following contributions:
\begin{itemize}
\item We devise a matched-filter quantized VB (MF-QVB) detection method for few-bit MIMO systems with known CSI. Unlike the VB-based detection method in~\cite{Thoota2021Variational} that assumes a known noise variance, the proposed MF-QVB method floats the noise variance as a latent variable and uses it to also account for residual inter-user interference. This latent variable is jointly estimated with the transmitted data symbol vector.

\item We develop a linear minimum mean-squared error quantized VB (LMMSE-QVB) detector that treats the noise covariance matrix as a latent variable, rather than simply assuming the noise covariance is a scaled identity matrix. The LMMSE-QVB detector offers performance similar to MF-QVB for independent and identically distributed (i.i.d.) channels, but significantly outperforms MF-QVB for spatially correlated channels.

\item We study the JED problem for few-bit MIMO systems and develop two methods, referred to as MF-QVB-JED and LMMSE-QVB-JED. The latter algorithm jointly estimates the channel matrix, the symbol data vectors, and the noise variances/covariance matrices. Again, this goes well beyond the prior work in \cite{Thoota2021Variational} that assumes a known scaled identity noise covariance.

\item We also present practical aspects of the VB framework to improve the implementation stability of the algorithms. We show via numerical results that the proposed VB detection algorithms provide much lower symbol error rates (SERs) compared to the conventional VB-based methods in~\cite{Thoota2021Variational}. The proposed QVB-JED algorithms also outperform FBM-DetNet in~\cite{Ly2022TWC}, particularly for spatially correlated channels.
\end{itemize}

The rest of this paper is organized as follows. We present the system model and the problem of interest in Section~\ref{sec:system_model}. Next, a brief introduction to the VB inference framework is given in Section~\ref{sec:VB_background}. Then, in Section~\ref{sec_VB_data_detection}, we derive the VB-based data detection method when the CSI is known. Section~\ref{sec_VB_joint_CE_DD} proposes the VB-based JED methods. We present practical implementation aspects of the VB framework as well as numerical results in Section~\ref{sec:simulation_results}. Finally, Section~\ref{sec:conclusion} concludes the paper.

\emph{Notation:} Scalars $x_{ij}$ and $[\mb{X}]_{ij}$ both denote the element at the $i$th row and $j$th column of a matrix $\mb{X}$; vector $\mb{x}_i$ denotes the $i$th column of a matrix $\mb{X}$; the operators $\tr\{\mb{X}\}$ and $|\mb{X}|$ represent the trace and determinant of a square matrix $\mb{X}$, respectively; the Frobenius norm of a matrix $\mb{X}$ is represented by $\|\mb{X}\|_F$; the distribution of a $K$-element complex Gaussian random vector with mean $\bs{\mu}$ and covariance matrix $\bs{\Sigma}$ is denoted by
$\mc{CN}(\mb{x};\bs{\mu},\bs{\Sigma}) = \frac{1}{\pi^K|\bs{\Sigma}|}\,\mr{exp}\big(-(\mb{x}-\bs{\mu})^H\bs{\Sigma}^{-1}(\mb{x}-\bs{\mu})\big)$, and is also written as $\mb{x}\sim \mc{CN}(\bs{\mu},\bs{\Sigma})$; the functions $\phi(x)$ and $\Phi(x)$ denote the PDF and cumulative distribution function (CDF) of a standard Gaussian random variable $\mc{N}(0,1)$; the operators $\mathbb{E}_{p(x)}[x]$ and $\mr{Var}_{p(x)}[x]$ denote the mean and variance of $x$ w.r.t. its distribution $p(x)$; in addition, we use $\lr{x}$, $\tau_x$, and $\lr{|x|^2} = |\lr{x}|^2 + \tau_x$ to denote the mean, variance, and second moment of $x$ w.r.t. a variational distribution $q(x)$. The symbols $\sim$ and $\propto$ indicate ``distributed according to'' and ``proportional to'', respectively. Finally, $\mathds{1}(\cdot)$ denotes the indicator function which equals one if the argument holds true, or zero otherwise.
	
	\section{System Model and Problem Statement}
	\label{sec:system_model}
	We consider an uplink massive MIMO system with $K$ single-antenna users and an $M$-antenna base station (BS). The symbols transmitted by the users are collected in the vector $\mb{x} = [x_1, \ldots,x_K]^T$, where $x_i$ corresponds to user-$i$ and is drawn from a discrete constellation $\mc{S}$, e.g., quadrature amplitude modulation (QAM) or phase-shift keying (PSK). The prior distribution of $x_i$ is
	\begin{eqnarray}\label{x-distribution}
		p(x_i) = \sum_{a\in\mc{S}} p_a\delta(x_i-a),
	\end{eqnarray}
	where $p_a$ corresponds to  a known prior probability of the constellation point $a \in \mc{S}$. 
	It is assumed that the symbols in $\mb{x}$ are independent of each other, i.e., $p(\mb{x}) = \prod_{i=1}^K p(x_i)$.
	
	Given $\mb{H} \in \mathbb{C}^{M\times K }$ as the uplink channel, the linear uplink MIMO system can be modeled as
	\begin{eqnarray}\label{system-model}
		\mb{r} = \mb{H}\mb{x} + \mb{n},
	\end{eqnarray}
	where $\mb{r}$ is the unquantized received signal vector and $\mb{n} \sim \mc{CN}( \mb{0},N_0\mb{I}_M)$ models the independent and identically distributed (i.i.d.) additive white Gaussian noise at the receiver. The channel vector $\mb{h}_i$ from user-$i$ to the BS is assumed to be distributed as $p(\mb{h}_i) = \mc{CN}(\mb{h}_i;\mb{0},\mb{C}_i)$ where $\mb{C}_i\triangleq \mathbb{E}[\mb{h}_i\mb{h}_i^H]$ is the covariance matrix that describes the spatial correlation between the receive antennas. 
	Finally, we assume that $\mathbb{E}[\mb{h}_i\mb{h}_j^H] = \mb{0}$, if $i\neq j$.
	
	We consider a block fading channel where a pilot matrix $\Xp$ followed by a data matrix $\Xd$ are transmitted in each block-fading interval. The unquantized received signals are given by
	\begin{eqnarray}\label{unquantized-model}
		\mb{R}_{\mr{p}} &=& \mb{H}\mb{X}_{\mr{p}} + \mb{N}_{\mr{p}} \nonumber \\
		\mb{R}_{\mr{d}} &=& \mb{H}\mb{X}_{\mr{d}} + \mb{N}_{\mr{d}},
	\end{eqnarray}
	where $\mb{X}_{\mr{p}} = \big[\mb{x}_{\mr{p},1},\ldots, \mb{x}_{\mr{p},T_{\mr{p}}}\big] \in \mathbb{C}^{K\times T_{\mr{p}}}$ with $\mb{x}_{\mr{p},t} = [x_{\mr{p},1t}, \ldots, x_{\mr{p},Kt}]^T$ and 
	$\mb{X}_{\mr{d}} = \big[\mb{x}_{\mr{d},1},\ldots, \mb{x}_{\mr{d},T_{\mr{d}}}\big] \in \mathbb{C}^{K\times T_{\mr{d}}}$ with $\mb{x}_{\mr{d},t} = [x_{\mr{d},1t}, \ldots, x_{\mr{d},Kt}]^T$. We assume that user-$i$ transmits with power $\rho_i$ during the pilot transmission phase.
	
	Each received analog signal is then quantized by a pair of $b$-bit ADCs to produce the quantized received signal:
    \begin{align} \label{quantized-model}
    \Re\{\mb{Y}_{\mr{p}}\} &=  \mc{Q}_b\left(\Re\{\mb{R}_{\mr{p}}\}\right), \;\; \Im\{\mb{Y}_{\mr{p}}\} =  \mc{Q}_b\left(\Im\{\mb{R}_{\mr{p}}\}\right); \nonumber \\
    \Re\{\mb{Y}_{\mr{d}}\} &=  \mc{Q}_b\left(\Re\{\mb{R}_{\mr{d}}\}\right), \;\; \Im\{\mb{Y}_{\mr{d}}\} =  \mc{Q}_b\left(\Im\{\mb{R}_{\mr{d}}\}\right),
    \end{align}
    where $\mc{Q}_b(\cdot)$ denotes the $b$-bit ADC operation which is applied separately to every element of its matrix or vector argument. It is assumed that $\mc{Q}_b(\cdot)$ performs $b$-bit uniform scalar quantization, which is characterized by a set of $2^b-1$ thresholds denoted
    as $\{d_1,\ldots,d_{2^b-1}\}$. Without loss of generality, we assume
    $-\infty = d_0 < d_1 <\ldots< d_{2^b-1} < d_{2^b} = \infty$. For a quantization step size of $\Delta$, the quantization thresholds are given by
    \begin{equation}
        d_{k} = (-2^{b-1}+k)\Delta, \; \text{for}\; k \in \mc{K}=\{1,\ldots,2^b-1\}.
    \end{equation}
    The quantized output $q$ is then defined as
    \begin{equation}
        q = \mc{Q}_b(r) =
        \begin{cases}
        d_{k} - \frac{\Delta}{2}, & \text{if}\; r\in(d_{k-1},d_k]\;\text{with}\;k\in\mc{K}\\
        (2^b-1)\frac{\Delta}{2},&\text{if}\;r\in(d_{2^b-1},d_{2^b}].
        \end{cases}
    \end{equation}
    We also define $q^\low=d_{k-1}$ and $q^\up=d_k$ as lower and upper thresholds of the quantization bin to which $q$ belongs.
    
    In this paper, we first study the data detection problem with known CSI, i.e., where the problem of interest is to detect the data matrix $\Xd$ using the received signal matrix $\Yd$ and the channel matrix $\mb{H}$. Then, in the following section we study the problem of joint channel estimation and data detection where $\mb{H}$ is estimated and $\Xd$ detected using knowledge of the pilot matrix $\Xp$ and the received signal matrices $\Yp$ and $\Yd$.

	\section{Background on Variational Bayes Inference}
	\label{sec:VB_background}
	This section presents a brief background on the VB method for approximate inference that will be developed for solving the problems of interest in this paper. In variational inference, the posterior distribution $p(\mb{x}|\mb{y})$ over a set of latent variables $\mb{x}$ given some observed data $\mb{y}$ is approximated by a variational distribution $q(\mb{x})$. A set of variational parameters describing $q(\mb{x})$ within a family of densities $\mc{Q}$ are determined to minimize the Kullback-Leibler (KL) divergence from $q(\mb{x})$ to $p(\mb{x}|\mb{y})$ \cite{Bishop-2006,Wainwright-2008}, i.e.,
	\begin{align}
		q(\mb{x}) &= \arg\min_{q(\mb{x}) \in \mc{Q}}\; \mr{KL}\big(q(\mb{x}) \|p(\mb{x}|\mb{y}) \big). 
	\end{align}
	The KL divergence is defined as
	\begin{align}
		\mr{KL}\big(q(\mb{x}) \|p(\mb{x}|\mb{y}) \big) &= \E_{q(\mb{x})} [\ln q(\mb{x}) ] - \E_{q(\mb{x})} [\ln p(\mb{x}|\mb{y})] \nonumber \\
		& = \E_{q(\mb{x})} \big[\ln q(\mb{x}) \big] - \E_{q(\mb{x})} \big[\ln p(\mb{x},\mb{y})\big] \nonumber \\
		&\quad + \ln p(\mb{y}). 
	\end{align}
	Since $p(\mb{y})$ is a constant w.r.t. $q(\mb{x})$, maximizing the evidence lower bound ($\mr{ELBO}$), 
	defined as
	\begin{eqnarray}
		\mr{ELBO}(q) =  \E_{q(\mb{x})} \big[\ln p(\mb{x},\mb{y})\big] - \E_{q(\mb{x})} \big[\ln q(\mb{x}) \big],
	\end{eqnarray}
	is equivalent to minimizing the KL divergence. 
	
	The maximum of $\mr{ELBO}(q)$ occurs when $q(\mb{x}) = p(\mb{x}|\mb{y})$. Since calculating the true posterior is often intractable, it is more convenient to consider a restricted family of distributions for $q(\mb{x})$. Here, the VB method assumes the \emph{mean field variational family}, such that
	\begin{eqnarray}\label{mean-field}
		q(\mb{x}) = \prod_{i=1}^m q_i(x_i).
	\end{eqnarray}
	In this mean field family, the latent variables are mutually independent and each is governed by a distinct factor in the variational distribution. 
	The general expression for the optimal solution $q_i(x_i)$ can be obtained as \cite{Bishop-2006}
	\begin{eqnarray}
		q_i(x_i) \propto \mr{exp}\left\{\big\langle{\ln p (\mb{y}|\mb{x}) + \ln p(\mb{x})\big\rangle_{-x_i}}\right\}.
	\end{eqnarray}
	Here, $\lr{\cdot}_{-x_i}$ denotes the expectation w.r.t. all latent variables except $x_i$ using the variational distribution $q_{-i}(\mb{x}_{-i}) = \prod_{j\neq i} q_{j}(x_{j})$. In the following, if $\lr{\cdot}$ is used, it means the variational expectation is taken w.r.t. all the latent variables in the argument. By iterating the update of $q_i(x_i)$ sequentially over all $i$, the $\mr{ELBO}(q)$ objective function can be monotonically improved. Thus, convergence to at least a local optimum of $\mr{ELBO}(q)$ is guaranteed \cite{Bishop-2006,Wainwright-2008}.
	
	In the following, we present a theorem on the variational posterior mean of multiple random variables that will be applied repeatedly later in the paper. 
		\begin{theorem}\label{theorem-1}
		Let $\mb{A}$, $\mb{y}$, and $\mb{x}$ of size $m\times n$, $m\times 1$, and $n \times 1$ be three independent random matrices (vectors) w.r.t. a variational distribution $q_{\mb{A},\mb{y}, \mb{x}}(\mb{A},\mb{y},\mb{x}) = q(\mb{A})q(\mb{y})q(\mb{x})$.  It is assumed that $\mb{A}$ is column-wise independent and let $\lr{\mb{a}_i}$ and $\bs{\Sigma}_{\mb{a}_i}$ be the variational mean and covariance matrix of the $i$th column of $\mb{A}$. Let $\lr{\mb{x}}$ and $\bs{\Sigma}_{\mb{x}}$ (and $\lr{\mb{y}}$ and $\bs{\Sigma}_{\mb{y}}$) be the variational mean and covariance matrix of $\mb{x}$ (and $\mb{y}$), respectively. For an arbitrary Hermitian matrix $\mb{B}$, let $\blr{(\mb{y}-\mb{A}\mb{x})^H\mb{B}(\mb{y}-\mb{A}\mb{x})}$ be the expectation of $(\mb{y}-\mb{A}\mb{x})^H\mb{B}(\mb{y}-\mb{A}\mb{x})$ w.r.t. $q_{\mb{A},\mb{y},\mb{x}}(\mb{A},\mb{y},\mb{x})$. We have
		\begin{align}\label{f-ABC}
		&	\blr{(\mb{y}-\mb{A}\mb{x})^H\mb{B}(\mb{y}-\mb{A}\mb{x})} \nonumber \\
			&= \big(\lr{\mb{y}}-\lr{\mb{A}}\lr{\mb{x}}\big)^H\mb{B}\big(\lr{\mb{y}}-\lr{\mb{A}}\lr{\mb{x}}\big) + \tr\{\mb{B}\bs{\Sigma_{\mb{y}}}\} \nonumber \\
			&\quad + \lr{\mb{x}}^H\mb{D}\lr{\mb{x}}  + \tr\big\{\bs{\Sigma}_{\mb{x}}\mb{D}\big\}+ \tr\big\{\bs{\Sigma}_{\mb{x}}\lr{\mb{A}^H}\mb{B}\lr{\mb{A}} \big\},
		\end{align}
		where $\mb{D} = \mr{diag}\big(\tr\{\mb{B}\bs{\Sigma}_{\mb{a}_1}\},\ldots, \tr\{\mb{B}\bs{\Sigma}_{\mb{a}_n}\}\big)$.
	\end{theorem}
	\begin{IEEEproof}
		The proof of this theorem is similar to the proof of Theorem~1 in~\cite{Duy-TSP-22}, except that $\mb{y}$ is now a random vector. Details of the proof are given in Appendix \ref{append}. 
	\end{IEEEproof}
	We note that if any of $\mb{A}$, $\mb{y}$, and $\mb{x}$ is deterministic, the corresponding covariance matrices $\{\bs{\Sigma}_{\mb{a}_i}\}$, $\bs{\Sigma}_{\mb{y}}$, and $\bs{\Sigma}_{\mb{x}}$ will be set to $\mb{0}$ and the expectation of $(\mb{y}-\mb{Ax})^H\mb{B}(\mb{y}-\mb{Ax})$ given in \eqref{f-ABC} can be simplified accordingly.


	
		\section{VB for Data Detection in Few-Bit MIMO Systems With CSIR}
	\label{sec_VB_data_detection}
	In this section, we develop new VB-based algorithms for solving the data detection problem in few-bit MIMO systems with known channel $\mb{H}$. For ease of presentation, we drop the subscripts $\mr{d},t$ indicating the data transmission at time index~$t$.
	
	\subsection{Proposed MF-QVB For Few-Bit MIMO Detection}
    The VB-based methods proposed in~\cite{Thoota2021Variational} assume prior information about the noise variance $N_0$. However,
    in practice, $N_0$ is not known \emph{a priori} and may need to be estimated. Furthermore, using the known noise variance, the conventional VB methods in~\cite{Thoota2021Variational} do not take into account the residual inter-user interference. Here, we consider the residual interference-plus-noise as an unknown parameter $N_0^{\mr{post}}$, which is postulated by the estimation in the VB framework~\cite{Duy-TSP-22}. For ease of computation, we use $\gamma = 1/N_0^{\mr{post}}$ to denote the precision to be estimated. 
    
	The joint distribution $p(\mb{y}, \mb{r}, \mb{x}; {\gamma}, {\mb{H}})$ of the observed variable $\mb{y}$ and the latent variables $\mb{r}$ and $\mb{x}$ can be factored as 
	\begin{align}   \label{MF-QVB-conditional}              
	p(\mb{y}, \mb{r}, \mb{x}; {\gamma}, {\mb{H}}) &= p(\mb{y}|\mb{r})  p(\mb{r}|\mb{x}; {\gamma}, \mb{H})p(\mb{x}) \nonumber \\
	&= \left[\prod_{m=1}^M p(y_m|r_m)\right] p(\mb{r}|\mb{x}; {\gamma},  \mb{H}) \left[\prod_{i=1}^K p(x_i)\right],
	\end{align}
	where $p(y_m|r_m)=\mathds{1}\big(r_m\in[y_m^\low,y_m^\up]\big)$ and $ p(\mb{r}|\mb{x}; {\gamma}, \mb{H}) =\mc{CN}(\mb{r};\mb{Hx},{\gamma}^{-1}\mb{I}_M)$. We note that the random vector $\mb{r}$ is comprised of conditional independent elements due to the same noise variance being imposed on the $M$ receive antennas.
	
	In the \textbf{\textit{E-step}}, for a currently fixed estimate $\hat{\gamma}$ of $\gamma$, we aim to derive the mean field variational distribution $q(\mb{r},\mb{x})$ of $\mb{r}$ and $\mb{x}$ given $\mb{y}$ such that
	\begin{align} 
	p(\mb{r}, \mb{x}|\mb{y}; \hat{\gamma}, \mb{H})  &\approx q(\mb{r}, \mb{x}) =  q(\mb{r})\left[\prod_{i=1}^K q(x_i)\right].
	\end{align}
	
	\textit{1) Updating $\mb{r}$.}  The variational distribution $q(\mb{r})$ is obtained by taking the expectation of the conditional in \eqref{MF-QVB-conditional} w.r.t. $q(\mb{x})$: 
	\begin{align}\label{r-dist}
\!\!\!	 q(\mb{r}) &\propto \mr{exp}\Big\{\blr{\ln p(\mb{y}|\mb{r}) + \ln p(\mb{r}|\mb{x}; \hat{\gamma}, \mb{H})}_{-\mb{r}}\Big\} \nonumber \\
	&\propto  \mr{exp}\Big\{\blr{\ln \mathds{1}\big(\mb{r}\in[\mb{y}^\low,\mb{y}^\up]\big)- \hat{\gamma}\|\mb{r} - \mb{H}\mb{x}\|^2}_{-\mb{r}}\Big\} \nonumber \\
	&\propto  \mathds{1}\big(\mb{r}\in[\mb{y}^\low,\mb{y}^\up]\big)\times  \mr{exp}\big\{\!-\hat{\gamma}\|\mb{r}-\mb{H}\lr{\mb{x}}\|^2\big\}\nonumber \\
	&\propto  \mathds{1}\big(\mb{r}\in[\mb{y}^\low,\mb{y}^\up]\big)\times \mc{CN}\big(\mb{r};\mb{H}\lr{\mb{x}},\hat{\gamma}^{-1}\mb{I}_M\big).
	\end{align}
	We note that variational distribution $q(\mb{r})$ is inherently separable as $\prod_{m=1}^M q(r_m)$ 
	without enforcing the mean field approximation on $q(\mb{r})$. Thus, the variational mean and variance can be obtained concurrently for all the elements of $\mb{r}$.  We see in \eqref{r-dist} that  $q(r_m)$ is  the truncated complex normal distribution obtained from bounding $r_m\sim \mc{CN}\big(s_m, \hat{\gamma}^{-1}\big)$, where $s_m = \mb{H}_{m,:}\lr{\mb{x}}$, to the interval $(y_m^\low, y_m^\up)$. Thus, its mean $\lr{r_m}$ and variance $\tau_{r_m}$ are given by  $\ms{F} _r\big(s_m,\hat{\gamma}, y_m^\low,y_m^\up\big)$ and $\ms{G} _r\big(s_m,\hat{\gamma}, y_m^\low,y_m^\up\big)$, respectively.\footnote{The computations of the mean $\ms{F} _r(\mu,\gamma,a,b)$ and variance $\ms{G} _r(\mu,\gamma,a,b)$  of an arbitrary complex normal distribution $\mc{CN}(\mu,\gamma^{-1})$ truncated to an interval $(a,b)$ are presented in Appendix \ref{append-A}.}
	
	 \begin{algorithm}[t]
	\caption{-- MF-QVB for Few-Bit MIMO Detection}
	\begin{algorithmic}[1]
		\State \textbf{Input:} $\mb{y}$, $\mb{H}$
		\State \textbf{Output:} $\hat{\mb{x}}$ 
		\State Initialize $\hat{x}_i^1 = 0$ and $\tau_{x_i}^1 = \mr{Var}_{p(x_i)}[x_i], \forall i$, $\hat{r}_m^1 = y_m$ and $\tau_{r_m}^1 = 0,\forall m$, and $\mb{e} = \hat{\mb{r}}^1- \mb{H}\hat{\mb{x}}^1$
		\For{$\ell=1,2,\ldots$}
    		\State $\hat{\gamma}^\ell \gets {M}/{\big(\|\mb{e}\|^2+\sum_{m=1}^M\tau^{\ell}_{r_m}+\sum_{i=1}^M\tau^{\ell}_{x_i}\|\mb{h}_i\|^2\big)}$
    	    \State $\hphantom{\hat{\gamma}^\ell}\mathllap{\mb{s}^\ell} \gets \hat{\mb{r}}^\ell - \mb{e}$
    	    \State $\hphantom{\hat{\gamma}^\ell}\mathllap{\hat{\mb{r}}^{\ell+1}} \gets \ms{F}_r\big(\mb{s}^\ell, \hat{\gamma}^\ell, \mb{y}^{\low}, \mb{y}^\up\big)$ \Comment{Update $\mb{r}$}
    		\State $\hphantom{\hat{\gamma}^\ell}\mathllap{\bs{\tau}_{\mb{r}}^{\ell+1}} \gets \ms{G}_r\big(\mb{s}^\ell ,\hat{\gamma}^\ell,\mb{y}^{\low}, \mb{y}^\up\big)$
    		\State $\hphantom{\hat{\gamma}^\ell}\mathllap{\mb{e}} \gets \mb{e} -\hat{\mb{r}}^{\ell} + \hat{\mb{r}}^{\ell+1}$ \Comment{Update residual}
    		\For{$i=1,\ldots,K$} \Comment{Update $\mb{x}$}
    			\State	$z_i^\ell \gets \hat{x}_i^\ell + \mb{h}_i^H\mb{e}/\|\mb{h}_i\|^2$ \Comment{Matched filter}
    			\State $\hphantom{z_i^\ell}\mathllap{\hat{x}_i^{\ell+1}} \gets \ms{F}_x\big(z_i^\ell, \hat{\gamma}^\ell\|\mb{h}_i\|^2\big)$
    			\State	$\hphantom{z_i^\ell}\mathllap{\tau_{x_i}^{\ell+1}} \gets \ms{G}_x\big(z_i^\ell, \hat{\gamma}^\ell\|\mb{h}_i\|^2\big)$
    			\State $\hphantom{z_i^\ell}\mathllap{\mb{e}} \gets \mb{e} + \mb{h}_i(\hat{x}_i^\ell - \hat{x}_i^{\ell+1})$ \Comment{Update residual}
    		\EndFor
		\EndFor
		\State $\forall i: \hat{x}_i \gets \argmax_{a\in\mc{S}} p_a\mc{CN}\big(z_i^\ell;a, 1/(\hat{\gamma}^\ell\|\mb{h}_i\|^2)\big)$.
		\end{algorithmic}
		\label{algo-1}
	\end{algorithm}

	\textit{2) Updating $x_i$.}  The variational distribution $q(x_i)$ is obtained by taking the expectation of the conditional in \eqref{MF-QVB-conditional} w.r.t. $q(\mb{r})\prod_{j\neq i}q(x_j)$: 
	\begin{align}\label{x-i-MF}
	q(x_i) &\propto \mr{exp}\big\{\big\langle \ln p(\mb{r}|\mb{x}; \hat{\gamma}, \mb{H}) + \ln p(x_i)\big\rangle_{-x_i}\big\} \nonumber \\
	&\propto p(x_i)\, \mr{exp}\big\{\!-\hat{\gamma}\blr{\|\mb{r}-\mb{H}\mb{x}\|^2}_{-x_i}\big\} \nonumber \\
	&\propto  p(x_i)\, \mr{exp}\bigg\{\!\!-\!\hat{\gamma}\bigg[\|\mb{h}_i\|^2 | x_i|^2  \nonumber \\
	&\quad\quad \quad\quad\quad\quad\quad\quad - 2\,\Re\bigg\{\mb{h}_i^H\!\bigg(\!\lr{\mb{r}} -\! \sum_{j\neq i}^K \!\mb{h}_j \lr{x_j}\!\bigg)x_i^*\!\bigg\}\bigg]\!  \bigg\} \nonumber \\
	&\propto  p(x_i)\,\exp \big\{\!-\! \hat{\gamma}\|{\mb{h}_i}\|^2 \big(|x_i|^2 - 2\,\Re\{x_i^*z_i\} \big) \big\} \nonumber \\
	&\propto p(x_i)\,\exp \big\{\!-\! \hat{\gamma}\|{\mb{h}_i}\|^2 |x_i - z_i|^2  \big\} \nonumber \\
	&\propto p(x_i)\,\mc{CN}\big(z_i;x_i,1/(\hat{\gamma}\|\mb{h}\|^2)\big) , 
	\end{align}
    where we define
	\begin{align} \label{z-i}
	z_i &=  \frac{\mb{h}_i^H}{\|{\mb{h}_i}\|^2}\Big(\lr{\mb{r}} - \sum_{j\neq i}^K \mb{h}_j\lr{x_j}\Big) \nonumber \\
	&= \lr{x_i} +  \frac{\mb{h}_i^H}{\|{\mb{h}_i}\|^2}\big(\lr{\mb{r}} - \mb{H}\lr{\mb{x}}\big)
	\end{align}
	with $\lr{x_i}$ being the \textit{currently fixed} nonlinear estimate of $x_i, \forall i$. We can see in \eqref{x-i-MF} that the mean field VB approximation decouples the few-bit MIMO system into an AWGN channel $z_i = x_i + \mc{CN}\big(0, 1/(\hat{\gamma}\|\mb{h}\|^2)\big)$ for user-$i$. The variational distribution $q(x_i)$ can be realized by normalizing $p(x_i)\,\mc{CN}\big(z_i;x_i,1/(\hat{\gamma}\|\mb{h}\|^2)\big)$.  The variational mean $\lr{x_i}$ and variance $\tau_{x_i}$ are now updated as $\ms{F}_x\big(z_i,\hat{\gamma}\|\mb{h}_i\|^2\big)$	and $\ms{G}_x\big(z_i,\hat{\gamma}\|\mb{h}_i\|^2\big)$, respectively.\footnote{The computations of the mean $\ms{F}_x(z,\gamma)$ and variance  $\ms{G}_x(z,\gamma)$ of a discrete random variable $x$ given a prior distribution $p(x)$ and the observation $z = x + \mc{CN}(0,\gamma^{-1})$ are presented in Appendix \ref{append-B}.}

	In the \textbf{\textit{M-step}}, the estimate of $\gamma$ is updated to maximize $\ln p(\mb{y},\mb{r},\mb{x};\gamma,\mb{H})$ w.r.t. $q(\mb{r},\mb{x})$, i.e.,
	\begin{align}
	   \hat{\gamma} &= \arg\max_\gamma\; \blr{\ln p (\mb{r}|\mb{x};\gamma;\mb{H})}\nonumber \\
	   &= \arg\max_\gamma\; M\ln\gamma - \gamma\blr{\|\mb{r}-\mb{Hx}\|^2}. 
	\end{align}
    Applying Theorem \ref{theorem-1} to evaluate the expectation $\blr{\|\mb{r} - \mb{H}\mb{x}\|^2}$, the new estimate of $\gamma$ is given by 
	\begin{align}\label{gamma}
	\!\!\!\hat{\gamma} = \frac{M}{\|\lr{\mb{r}}-\mb{H}\lr{\mb{x}}\|^2 + \sum_{m=1}^M\tau_{r_m}+ \sum_{i=1}^K \tau_{x_{i}}\|\mb{h}_i\|^2}.
	\end{align}
	
	By iteratively optimizing $q(\mb{r})$, $\{q(x_i)\}$, and updating $\hat{\gamma}$, we obtain the variational Bayes expectation-maximization (VBEM) algorithm for estimating $\mb{r}$, $\mb{x}$, and $\gamma$. Similar to our previous work \cite{Duy-TSP-22}, we refer this scheme to as the \textit{\textbf{MF-QVB algorithm}} due to the use of the matched-filter $\mb{h}_i^H/\|\mb{h}_i\|^2$ to obtain the linear estimate $z_i$ of $x_i$ in \eqref{z-i}. If $\gamma$ is fixed to $N_0^{-1}$, the \textit{\textbf{MF-QVB algorithm}} will be referred to as the \textit{\textbf{conv-QVB algorithm}}, that was investigated as the QVB-CSIR algorithm in \cite{Thoota2021Variational}. The MF-QVB approach is summarized in Algorithm~\ref{algo-1}.  Here, we use $\hat{\mb{r}}^{\ell}$, $\hat{x}_i^{\ell}$, and $\hat{\gamma}^{\ell}$ to replace $\lr{\mb{r}}$, $\lr{x_i}$, and $\hat{\gamma}$ at iteration $\ell$ and each iteration consists of one round of updating the estimation of $\mb{r}$, $\mb{x}$, and $\gamma$. To reduce the complexity of the algorithm, we include the residual term $\mb{e}$, which is initialized as $\hat{\mb{r}}^{1}-\mb{H}\hat{\mb{x}}^1$. We also define $\mb{s}^\ell = \hat{\mb{r}}^{\ell} - \mb{e}$ as an efficient way to compute $\mb{H}\hat{\mb{x}}^\ell$ at iteration-$\ell$. The computation of $\mb{r}^{\ell+1}$ (and $\bs{\tau}_{\mb{r}}^{\ell+1}$) can be carried out element-wise in parallel.  
	
   \vspace{-0.15cm}
	\subsection{Proposed LMMSE-QVB For Few-Bit MIMO Detection}
	We now develop the LMMSE-QVB method for few-bit MIMO detection that uses a postulated noise covariance matrix $\mb{C}^{\mr{post}}$ instead of the postulated noise variance $N_0^{\mr{post}}$ in the MF-QVB method. The idea of using a postulated noise covariance matrix $\mb{C}^{\mr{post}}$ was proposed in~\cite{Duy-TSP-22} but for infinite-resolution ADCs. For ease of computation, we use $\bs{\Gamma} = (\mb{C}^{\mr{post}})^{-1}$ as the precision matrix to be estimated. 
 
    The joint distribution $p(\mb{y}, \mb{r}, \mb{x}; \bs{\Gamma}, {\mb{H}})$ of the observed variable $\mb{y}$ and the latent variables $\mb{r}$ and $\mb{x}$ at time slot $t$ can be factored as
    \vspace{-0.15cm}
	\begin{align}   \label{MMSE-VB-conditional}              
	&p(\mb{y}, \mb{r}, \mb{x}; \bs{\Gamma}, {\mb{H}}) \nonumber \\
	& =p(\mb{y}|\mb{r})  p(\mb{r}|\mb{x}; \bs{\Gamma}, \mb{H})p(\mb{x}) \nonumber \\
	&= \left[\prod_{m=1}^M p(y_m|r_m)\right] p(\mb{r}|\mb{x}; \bs{\Gamma}, \mb{H}) \left[\prod_{i=1}^K p(x_i)\right],
	\end{align}
	where $p(\mb{r}|\mb{x},\bs{\Gamma};\mb{H}) =\mc{CN}\big(\mb{r};\mb{Hx},\bs{\Gamma}^{-1}\big)$. We note that the random vector $\mb{r}$ is no longer comprised of conditional independent elements, since the noise covariance matrix $\bs{\Gamma}^{-1}$ is in general non-diagonal.
	
        In the \textbf{\textit{E-step}}, for a currently fixed estimate $\hat{\bs{\Gamma}}$ of $\bs{\Gamma}$, we aim to derive the mean field variational distribution $q(\mb{r},\mb{x})$ of $\mb{r}$ and $\mb{x}$ given $\mb{y}$ such that
        \vspace{-0.1cm}
	\begin{align} 
	\!\!\!p(\mb{r}, \mb{x}|\mb{y};\hat{\bs{\Gamma}},\mb{H})  &\approx q(\mb{r}, \mb{x}) = \left[\prod_{m=1}^M q(r_m)\right]\!\left[\prod_{i=1}^K q(x_i)\right]\!.
	\end{align}

	\textit{1) Updating $r_m$.}  The variational distribution $q(r_m)$ is obtained by taking the expectation of the conditional in \eqref{MMSE-VB-conditional} w.r.t. $q(\mb{x})\prod_{n\neq m} q(r_n)$: 
	\begin{align}\label{r-m}
	q(r_m) &\propto \mr{exp}\Big\{\blr{ \ln p(y_m|r_m) + \ln p(\mb{r}|\mb{x};\hat{\bs{\Gamma}},\mb{H})}_{-r_m}\Big\} \nonumber \\
	&\propto  \mr{exp}\Big\{\ln \mathds{1}\big(r_m\in[y_m^\low,y_m^\up]\big) \nonumber \\
	&\quad\quad\quad\;\;\; - \blr{(\mb{r} - \mb{H}\mb{x})^H\hat{\bs{\Gamma}}(\mb{r} - \mb{H}\mb{x})}_{-r_m}\Big\} \nonumber \\
	&\propto  \mathds{1}\big(r_m\in[y_m^\low,y_m^\up]\big) \nonumber \\
	&\;\;\;\times \mr{exp}\Big\{ -\hat{\gamma}_{mm} \big(|r_m|^2-2\,\Re\{r_m^*\mb{H}_{m,:}\lr{\mb{x}}\}\big)^2 \nonumber \\
	&\quad\quad\quad\quad - 2\sum_{n\neq m}^M \Re\big\{r_m^*\hat{\gamma}_{mn}\big(\lr{r_n}-\mb{H}_{n,:}\lr{\mb{x}}\big)\big\}\Big\} \nonumber \\
	&\propto  \mathds{1}\big(r_m\!\in\![y_m^\low,y_m^\up]\big)\! \times \mr{exp}\big\{-\hat{\gamma}_{mm}|r_m - s_m|^2\big\} \nonumber \\
	&\propto  \mathds{1}\big(r_m\!\in\![y_m^\low,y_m^\up]\big) \!\times\mc{CN} \big(r_m;s_m,\hat{\gamma}_{mm}^{-1}\big),
	\end{align}
	where $s_m$ is now defined as
	\begin{align}
	s_m &= \mb{H}_{m,:}\lr{\mb{x}} - \hat{\gamma}_{mm}^{-1} \sum_{n\neq m}^M \hat{\gamma}_{mn}\big(\lr{r_n} - \mb{H}_{n,:}\lr{\mb{x}}\big) \nonumber \\
	&= \lr{r_m} - \frac{\hat{\bs{\Gamma}}_m^H}{\hat{\gamma}_{mm}}\big(\lr{\mb{r}} - \mb{H}\lr{\mb{x}}\big),
	\end{align}
	$\lr{r_m}$ is the \emph{currently fixed} nonlinear estimate of $r_m$ and $\hat{\bs{\Gamma}}_m$ is the $m$th column of the Hermitian matrix $\hat{\bs{\Gamma}}$. We can see in \eqref{r-m} that the variational distribution $q(r_m)$ is the truncated  complex normal distribution obtained from bounding $r_m\sim \mc{CN}(s_m,\hat{\gamma}_{mm}^{-1})$ to the interval $(y_m^\low, y_m^\up)$.  Thus, its mean $\lr{r_m}$ and variance $\tau_{r_m}$ are updated as  $\ms{F} _r\big(s_m,\hat{\gamma}_{mm}, y_m^\low,y_m^\up\big)$ and $\ms{G}_r\big(s_m,\hat{\gamma}_{mm}, y_m^\low,y_m^\up\big)$, respectively.
	
	\textit{2) Updating $x_i$.}  The variational distribution $q(x_i)$ is obtained by taking the expectation of the conditional in \eqref{MF-QVB-conditional} w.r.t. $q(\mb{r})\prod_{j\neq i}q(x_j)$: 
	\begin{align}\label{x-i-MMSE}
	q(x_i) &\propto \mr{exp}\Big\{\big\langle \ln p(\mb{r}|\mb{x};\hat{\bs{\Gamma}},\mb{H}) + \ln p(x_i)\big\rangle_{-x_i}\Big\} \nonumber \\
	&\propto p(x_i)\, \mr{exp}\Big\{-\blr{(\mb{r}-\mb{Hx})^H\hat{\bs{\Gamma}}(\mb{r}-\mb{Hx})}_{-x_i}\Big\} \nonumber\\
	&\propto p(x_i)\,\mr{exp} \big\{- \mb{h}_i^H\hat{\bs{\Gamma}}\mb{h}_i |x_i - z_i|^2\big\} \nonumber \\
	&\propto p(x_i)\,\mc{CN}\big(z_i;x_i,{1}/{\big(\mb{h}_i^H\hat{\bs{\Gamma}}\mb{h}_i\big)} \big),
	\end{align}
	where $z_i$ is a linear estimate of $x_i$ that is now defined as
	\begin{eqnarray} \label{z-i-LMMSE}
	z_i &=& \frac{\mb{h}^H_i\hat{\bs{\Gamma}}}{\mb{h}_i^H\hat{\bs{\Gamma}}\mb{h}_i}\bigg(\lr{\mb{r}} - \sum_{j\neq i}^K \mb{h}_j\lr{x_j}\bigg) \nonumber \\
	&=& \lr{x_i} + \frac{\mb{h}^H_i\hat{\bs{\Gamma}}}{\mb{h}_i^H\hat{\bs{\Gamma}}\mb{h}_i}\big(\lr{\mb{r}} - \mb{H}\lr{\mb{x}}\big),
	\end{eqnarray}
	and $\lr{x_i}$ is the \emph{current} nonlinear estimate of $x_i$. 
	Here, $z_i$ is the LMMSE estimate of $x_i$ using the LMMSE filter ${\mb{h}^H_i\hat{\bs{\Gamma}}}/{(\mb{h}_i^H\hat{\bs{\Gamma}}\mb{h}_i)}$. The variational distribution $q(x_i)$ can be realized by normalizing $p(x_i)\,\mc{CN}\big(z_i;x_i,{1}/{(\mb{h}_i^H\hat{\bs{\Gamma}}\mb{h}_i)} \big)$. The variational mean $\lr{x_i}$ and variance $\tau_{x_i}$ are updated as $\ms{F}_x\big(z_i,\mb{h}_i^H\hat{\bs{\Gamma}}\mb{h}_i\big)$ and $\ms{G}_x\big(z_i,\mb{h}_i^H\hat{\bs{\Gamma}}\mb{h}_i\big)$, respectively.
	
	In the \textbf{\textit{M-step}}, the estimate of $\bs{\Gamma}$ is updated to maximize $\ln p(\mb{y},\mb{r},\mb{x};\bs{\Gamma},\mb{H})$ w.r.t. $q(\mb{r},\mb{x})$, i.e.,
	\begin{align}
	\hat{\bs{\Gamma}} &= \arg\max_{\bs{\Gamma}}\; \blr{\ln p(\mb{r}|\mb{x}; \bs{\Gamma},\mb{H})} \nonumber \\
	&=\arg\max_{\bs{\Gamma}}\; \ln|\bs{\Gamma}| - \blr{(\mb{r}-\mb{Hx})^H\bs{\Gamma}(\mb{r}-\mb{Hx})}.
	\end{align}
	By applying Theorem \ref{theorem-1}, we have 
	\begin{align} 
	&\blr{(\mb{r}-\mb{Hx})^H\bs{\Gamma}(\mb{r}-\mb{Hx})} \nonumber \\
	&=\tr\Big\{\Big[\big(\lr{\mb{r}}-\mb{H}\lr{\mb{x}}\big)\big(\lr{\mb{r}}-\mb{H}\lr{\mb{x}}\big)^H\! + \bs{\Sigma}_{\mb{r}} +\mb{H}\mb{\Sigma}_{\mb{x}}\mb{H}^H\Big]\bs{\Gamma}\Big\},
	\end{align}
	where $\bs{\Sigma}_{\mb{r}} = \mr{diag}(\tau_{r_1},\ldots,\tau_{r_M})$ and $\bs{\Sigma}_{\mb{x}} = \mr{diag}(\tau_{x_1},\ldots,\tau_{x_K})$. Thus, a new estimate of $\bs{\Gamma}$ is given by
	\begin{align}\label{W-EM}
	\hat{\bs{\Gamma}} = \left(\!\big(\lr{\mb{r}} - \mb{H}\lr{\mb{x}}\big)\big(\lr{\mb{r}} - \mb{H}\lr{\mb{x}}\big)^H\!\! + \bs{\Sigma}_{\mb{r}} + \mb{H}\bs{\Sigma}_{\mb{x}}\mb{H}^H\right)^{-1}\!.
	\end{align}
	We note that the matrix inversion in \eqref{W-EM} often results in numerical errors due to the rank deficiency of $\big(\lr{\mb{r}}-\mb{H}\lr{\mb{x}}\big)\big(\lr{\mb{r}}-\mb{H}\lr{\mb{x}}\big)^H+\bs{\Sigma}_{\mb{r}} +\mb{H}\mb{\Sigma}_{\mb{x}}\mb{H}^H$. Similar to the approach in \cite{Duy-TSP-22}, we propose to use the following estimator	
	\begin{align}\label{W}
	\hat{\bs{\Gamma}} = \left(\frac{\|\lr{\mb{r}} - \mb{H}\lr{\mb{x}}\|^2}{M}\mb{I}_M + \bs{\Sigma}_{\mb{r}} + \mb{H}\bs{\Sigma}_{\mb{x}}\mb{H}^H\right)^{-1}.
	\end{align}
	for the precision matrix $\bs{\Gamma}$.
	
	By iteratively optimizing $\{q(r_m)\}$, $\{q(x_i)\}$, and $\hat{\bs{\Gamma}}$, we obtain the VBEM algorithm for estimating $\mb{r}$, $\mb{x}$, and $\bs{\Gamma}$. We refer to this scheme as the \textit{\textbf{LMMSE-QVB algorithm}} due to the use of the LMMSE filter $\mb{h}_i^H\hat{\bs{\Gamma}}/(\mb{h}_i^H\hat{\bs{\Gamma}}\mb{h}_i)$ to obtain the linear estimate $z_i$ of $x_i$ in \eqref{z-i-LMMSE}. The LMMSE-QVB approach is summarized in Algorithm \ref{algo-2}. As before, we use $\hat{r}_m^{\ell}$, $\hat{x}_i^{\ell}$, and $\hat{\bs{\Gamma}}^{\ell}$ to replace $\lr{r_m}$, $\lr{x_i}$, and $\hat{\bs{\Gamma}}$ at iteration $\ell$ and each iteration consists of one round of updating the estimation of  $\mb{r}$, $\mb{x}$, and $\bs{\Gamma}$. Unlike MF-QVB, the LMMSE-QVB algorithm requires sequential updates over $\{\hat{r}_m^{\ell}\}$. Note that LMMSE-QVB is equivalent to MF-QVB in Algorithm \ref{algo-1} if $\hat{\bs{\Gamma}}^{\ell}$ is set as $\hat{\gamma}^{\ell}\mb{I}_M$.
	 
\begin{algorithm}[t]
	\caption{-- LMMSE-QVB for Few-Bit MIMO Detection}
	\begin{algorithmic}[1]
		\State \textbf{Input:} $\mb{y}$, $\mb{H}$
		\State \textbf{Output:} $\hat{\mb{x}}$ 
		\State Initialize $\hat{x}_i^1 = 0$ and $\tau_{x_i}^1 = \mr{Var}_{p(x_i)}[x_i], \forall i$, $\hat{r}_m^1 = y_m$ and $\tau_{r_m}^1 = 0,\forall m$, and $\mb{e} = \hat{\mb{r}}^1- \mb{H}\hat{\mb{x}}^1$
		\For{$\ell=1,2,\ldots$}
    		\State $\bs{\Sigma}_{\mb{r}} \gets \mr{diag}(\tau_{r_1}^\ell,\ldots,\tau_{r_M}^\ell)$
			\State $\hphantom{\bs{\Sigma}_{\mb{r}}}\mathllap{\bs{\Sigma}_{\mb{x}}} \gets \mr{diag}(\tau_{x_1}^\ell,\ldots,\tau_{x_K}^\ell)$
			\State $\hphantom{\bs{\Sigma}_{\mb{r}}}\mathllap{\hat{\bs{\Gamma}}^\ell} \leftarrow \big((\|\mb{e}\|^2/M)\mb{I}_M+ \bs{\Sigma}_{\mb{r}}  + \mb{H}\mb{\Sigma}_{\mb{x}}\mb{H}^H\big)^{-1}$
			\For{$m=1,\ldots,M$} \Comment{Update $\mb{r}$}
    	    \State $s_m^\ell \gets \hat{r}_m^\ell  - (\hat{\bs{\Gamma}}_{m}^{\ell})^H\mb{e}/w^\ell_{mm}$
  	        \State $\hphantom{s_m^\ell}\mathllap{\hat{r}_m^{\ell+1}} \gets \ms{F}_r\big(s_m^\ell,\hat{w}_{mm}^\ell,y_m^{\low}, y_m^\up\big)$
      	    \State $\hphantom{s_m^\ell}\mathllap{\tau_{r_m}^{\ell+1}} \gets \ms{G} _r\big(s_m^\ell,\hat{w}_{mm}^\ell,y_m^{\low}, y_m^\up\big)$
      	    \State $\hphantom{s_m^\ell}\mathllap{e_m} \leftarrow e_m- \hat{r}_m^\ell + \hat{r}_m^{\ell+1}$ \Comment{Update residual}
      	    \EndFor
    		\For{$i=1,\ldots,K$} \Comment{Update $\mb{x}$}
    			\State	$z_i^\ell \gets \hat{x}_i^\ell + \mb{h}_i^H\hat{\bs{\Gamma}}^\ell\mb{e}/(\mb{h}_i^H\hat{\bs{\Gamma}}^{\ell}\mb{h}_i)$ \Comment{LMMSE}
    			\State $\hphantom{z_i^\ell}\mathllap{\hat{x}_i^{\ell+1}} \gets \ms{F}_x\big(z_i^\ell, \mb{h}_i^H\hat{\bs{\Gamma}}^\ell\mb{h}_i\big)$
    			\State	$\hphantom{z_i^\ell}\mathllap{\tau_{x_i}^{\ell+1}} \gets \ms{G}_x\big(z_i^\ell, \mb{h}_i^H\hat{\bs{\Gamma}}^\ell\mb{h}_i\big)$
    			\State $\hphantom{z_i^\ell}\mathllap{\mb{e}} \gets \mb{e} + \mb{h}_i(\hat{x}_i^\ell - \hat{x}_i^{\ell+1})$ \Comment{Update residual}
    		\EndFor
		\EndFor
		\State $\forall i: \hat{x}_i \gets \argmax_{a\in\mc{S}}\, p_a\mc{CN}\big(z_i^\ell;a, 1/(\mb{h}_i^H\hat{\bs{\Gamma}}^{\ell}\mb{h}_i)\big)$.
		\end{algorithmic}
		\label{algo-2}
	\end{algorithm}

	
	\subsection{Practical Aspects of Implementing MF/LMMSE-QVB}
    \textit{1) Computing $\lr{r_m}$ and $\tau_{r_m}$:} In MF-QVB, $\lr{r_m}$ and $\tau_{r_m}$ are updated as $\ms{F} _r\big(s_m,\hat{\gamma}, y_m^\low,y_m^\up\big)$ and $\ms{G} _r\big(s_m,\hat{\gamma}, y_m^\low,y_m^\up\big)$, respectively. The computations of these two terms presented in Appendix \ref{append-A} can result in catastrophic cancellation when $\Phi(\beta) = \Phi(\alpha)$, even if $\beta = \sqrt{2\hat{\gamma}}(y_m^\low - s_m)$ and $\alpha = \sqrt{2\hat{\gamma}}(y_m^\up - s_m)$ are different. 
    This numerical error often occurs when $s_m$ is not inside the interval $(y_m^\low,y_m^\up)$ and $\lr{\gamma}$ is large, i.e., for high SNR. To improve the robustness of the computation, we use the logistic CDF $F(x;c) = \frac{1}{1 + e^{-cx}}$ and logistic PDF $p(x;c) = cF(x;c)(1-F(x;c))$ in lieu of the normal CDF $\Phi(x)$ and normal PDF $\phi(x)$. We choose $c = 3/\sqrt{\pi}$ to impose a unit variance on the logistic distribution. Note that $F(x)$ is much easier to compute than $\Phi(x)$. We have observed through numerous simulations that this modification eliminates numerical errors due to the heavier tails in the logistic distribution compared to the normal distribution. Interestingly, the detection accuracy is slightly better when using $F(x;c)$ and $p(x;c)$ rather than $\Phi(x)$ and $\phi(x)$, even when no numerical errors occur using the latter approach. We also use $F(x;c)$ and $p(x;c)$ in LMMSE-QVB and observe the same effect.
    
    \textit{2) Using $\mb{e}$:} The residual term $\mb{e}$ is included in MF/LMMSE-QVB reduce the computational complexity of these algorithms. Due to the sequential nature of VB, $s_m$ and $z_i$ are computed using the latest updated values of $\lr{\mb{r}}$ and $\lr{\mb{x}}$. Instead of computing $\lr{\mb{r}}-\mb{H}\lr{\mb{x}}$ each iteration, which induces a complexity of $\mc{O}(MK)$, we use the current value of the residual term $\mb{e}$. The update of $\mb{e}$ reflecting any update on estimation of $\mb{r}$ or $x_i$ only induces a complexity of $\mc{O}(M)$.
    

	\section{VB for Joint Channel Estimation and Data Detection in Few-Bit MIMO Systems}\label{sec_VB_joint_CE_DD}
    The algorithms in the previous section assumed that the CSI was already obtained by some other method prior to data detection. In this section, we generalize the MF-QVB and LMMSE-QVB approaches to perform joint channel estimation and data detection.

	\subsection{Proposed MF-QVB-JED Algorithm} \label{sec-MF-QVB-JED}
	We denote $\gp$ and $\bs{\gamma}_{\mr{d}} = [\gamma_{\mr{d},1},\ldots,\gamma_{\mr{d},\Td}]^T$ as the precision of the noise in the pilot transmission phase and the $\Td$ data transmission time slots, respectively. The factorization of the joint distribution of all the observed and latent variables in the system model \eqref{unquantized-model}--\eqref{quantized-model} is given in \eqref{CE-DD-conditional}, where $p(\rpt|\mb{H};\gamma_{\mr{p}}, \xpt) = \mc{CN}\big(\rpt;\mb{H}\xpt,\gp^{-1}\mb{I}_M\big)$ and $p(\rdt|\mb{H},\mb{x}_{\mr{d},t};\gdt) = \mc{CN}\big(\rdt;\mb{H}\xdt,\gdt^{-1}\mb{I}_M\big)$.
	\begin{figure*}
	\begin{align}\label{CE-DD-conditional}
		&p(\Yp,\Yd,\Rp,\Rd,\mb{X}_{\mr{d}},\mb{H};{\gamma}_{\mr{p}},\bs{\gamma}_{\mr{d}}, \mb{X}_{\mr{p}},\{\mb{C}_i\}) \nonumber \\
		&= p(\Yp|\Rp)p(\Rp|\mb{H};\gamma_{\mr{p}},\mb{X}_{\mr{p}}) p(\Yd|\Rd)p(\Rd|\mb{X}_{\mr{d}},\mb{H};\bs{\gamma}_{\mr{d}}) p(\mb{X}_{\mr{d}}) p(\mb{H}|\{\mb{C}_i\}) \nonumber \\
		& = \left[\prod_{t=1}^{\Tp} p(\ypt|\rpt)p(\rpt|\mb{H};\gamma_{\mr{p}}, \xpt) \right]\left[\prod_{t=1}^{\Td} p(\ydt|\rdt)p(\rdt|\mb{H},\xdt;\gdt) p(\xdt)\right]  \prod_{i=1}^Kp(\mb{h}_i|\mb{C}_i).
	\end{align}
	\vspace{-0.1cm}
	\hrule
	\vspace{-0.1cm}
	\setcounter{equation}{34}
		\begin{eqnarray}
			q(\mb{h}_i)  
			&\propto& \mr{exp}\bigg\{-\bigg\langle\hat{\gamma}_{\mr{p}}\sum_{t=1}^{T_{\mr{p}}}  \big\|\mb{r}_{\mr{p},t}-\mb{H}\mb{x}_{\mr{p},t}\big\|^2 + \sum_{t=1}^{T_{\mr{d}}} \hat{\gamma}_{\mr{d},t}\big\|\mb{r}_{\mr{d},t}-\mb{H}\mb{x}_{\mr{d},t}\big\|^2+ \mb{h}_i^H\mb{C}_i^{-1}\mb{h}_i\bigg\rangle_{-\mb{h}_i} \bigg\} \nonumber \\
			&\propto&\mr{exp}\bigg\{-\mb{h}_i^H\bigg[\bigg(\hat{\gamma}_{\mr{p}} \sum_{t=1}^{\Td} |\xpit|^2 + \sum_{t=1}^{T_{\mr{d}}} \hat{\gamma}_{\mr{d},t}\lr{|x_{\mr{d},it}|^2}\bigg)\mb{I}_M + \mb{C}_i^{-1}\bigg]\mb{h}_i \nonumber \\
			&&\quad\quad  + 2\,\Re\bigg\{\mb{h}_i^H \bigg[\hat{\gamma}_{\mr{p}}\sum_{t=1}^{\Tp} \!\bigg(\!\lr{\mb{r}_{\mr{p},t}}\! -\! \sum_{j\neq i}^K \langle\mb{h}_j\rangle \xpjt\! \bigg) \xpit^*  +\! \sum_{t=1}^{T_{\mr{d}}} \hat{\gamma}_{\mr{d},t} \Big(\lr{\mb{r}_{\mr{d},t}} - \!\sum_{j\neq i}^K \langle\mb{h}_j\rangle\langle x_{\mr{d},jt} \rangle \Big)\lr{x_{\mr{d},it}^*}\bigg]\! \bigg\}\! \bigg\}. \label{q-h-i-quantized} 
		\end{eqnarray} 
		\vspace{-0.1cm}
		\hrulefill
			\vspace{-0.1cm}
		\begin{eqnarray} 
			\bs{\Sigma}_{\mb{h}_i} &=& \bigg[\bigg(\hat{\gamma}_{\mr{p}}\sum_{t=1}^{\Tp} |\xpit|^2 + \sum_{t=1}^{T_{\mr{d}}} \hat{\gamma}_{\mr{d},t}\lr{|x_{\mr{d},it}|^2}\bigg)\mb{I}_M + \mb{C}_i^{-1}\bigg]^{-1} \label{Sigma-h-noise-quantized} \\
			\lr{\mb{h}_i} &=& \bs{\Sigma}_{\mb{h}_i}\bigg[\hat{\gamma}_{\mr{p}}\sum_{t=1}^{T_{\mr{p}}} \!\bigg(\!\lr{\mb{r}_{\mr{p},t}}\! -\! \sum_{j\neq i}^K \langle\mb{h}_j\rangle \xpjt\! \bigg) \xpit^*   +\! \sum_{t=1}^{T_{\mr{d}}}\hat{\gamma}_{\mr{d},t} \Big(\lr{\mb{r}_{\mr{d},t}} - \!\sum_{j\neq i}^K \langle\mb{h}_j\rangle\langle x_{\mr{d},jt} \rangle \Big)\lr{x_{\mr{d},it}^*}\bigg].  \label{mean-h-quantized}
		\end{eqnarray}
		\vspace{-0.2cm}
		\hrulefill
	\end{figure*}

In the \textbf{\textit{E-step}}, for currently fixed estimates $\hat{\gamma}_{\mr{p}}$ and $\hat{\bs{\gamma}}_{\mr{d}}$ of $\gp$ and $\bs{\gamma}_{\mr{d}}$, respectively, we aim to obtain the mean field variational distribution $q(\Rp,\Rd,\mb{H},\Xd)$ of $\Rp$, $\Rd$,  $\mb{H}$, and $\Xd$ given $\Yp$ and $\Yd$ such that
\setcounter{equation}{30}
	\begin{align}
		&p(\Rp,\Rd,\mb{X}_{\mr{d}},\mb{H}|\Yp,\Yd;\hat{\gamma}_{\mr{p}},\hat{\bs{\gamma}}_{\mr{d}},\Xp,\{\mb{C}_i\}) \nonumber \\
		&\approx q(\Rp,\Rd,\mb{X}_{\mr{d}},\mb{H}) \nonumber \\
		&= \Bigg[\prod_{t=1}^{T_{\mr{p}}}q(\rpt)\Bigg]\!\left[ \prod_{t=1}^{T_{\mr{d}}} q(\rdt)\right]\! \left[\prod_{i=1}^Kq(\mb{h}_i)\right]\! \left[\prod_{i=1}^K\prod_{t=1}^{T_{\mr{d}}} q(x_{\mr{d},it})\right].
	\end{align}
	
	\textit{1) Updating $\rpt$.} Taking the expectation of the conditional \eqref{CE-DD-conditional} w.r.t. all latent variables except $\rpt$, the variational distribution $q(\rpt)$ is given by 
	\begin{align}\label{r-p-t}
	& q(\rpt) \nonumber \\
	&\propto \mr{exp}\Big\{\big\langle \ln p(\ypt|\rpt) + \ln p(\rpt|\mb{H};\hat{\gamma}_{\mr{p}},\mb{x}_{\mr{p},t}) \big\rangle_{-\rpt}\Big\} \nonumber \\
	&\propto  \mr{exp}\Big\{\blr{\ln \mathds{1}\big(\rpt\in[\ypt^\low,\ypt^\up]\big)-\nonumber \\ &\hspace{4.2cm}\hat{\gamma}_{\mr{p}}\|\mb{r}_{\mr{p},t}-\mb{H}\mb{x}_{\mr{p},t}\|^2}_{-\rpt}\Big\} \nonumber \\
	&\propto \mathds{1}\big(\rpt\in[\ypt^\low,\ypt^\up]\big)\times \mr{exp}\Big\{\!-\hat{\gamma}_{\mr{p}}\|\mb{r}_{\mr{p},t}-\lr{\mb{H}}{\mb{x}_{\mr{p},t}}\|^2\Big\}.
	\end{align}
	We note that the variational distribution $q(\rpt)$ is inherently separable as $\prod_{m=1}^M q(\rpmt)$ and the variational distribution $q(\rpmt)$ is the complex complex normal distribution obtained from bounding $\rpmt \sim \mc{CN}\big(\lr{\mb{H}_{m,:}}\xpt,\hat{\gamma}_{\mr{p}}^{-1}\big)$ to the interval $(\ypmt^\low,\ypmt^\up)$. The variational mean $\lr{\rpmt}$ and variance $\tau_{\rpmt}$ are given by $\ms{F}_r\big(\lr{\mb{H}_{m,:}}\xpt,\hat{\gamma}_{\mr{p}}^{-1}, \ypmt^\low,\ypmt^\up\big)$ and $\ms{G}_r\big(\lr{\mb{H}_{m,:}}\xpt,\hat{\gamma}_{\mr{p}}^{-1}, \ypmt^\low,\ypmt^\up\big)$, respectively.
	
	\textit{2) Updating $\rdt$.} Taking the expectation of the conditional \eqref{CE-DD-conditional} w.r.t. all latent variables except $\rdt$, the variational distribution $q(\rpt)$ is given by 
	\begin{align}\label{r-d-t}
	& q(\rdt) \nonumber \\
	&\propto \mr{exp}\Big\{\big\langle \ln p(\ydt|\rdt) + \ln p(\rdt|\mb{H},\mb{x}_{\mr{d},t};\hat{\gamma}_{\mr{d},t}) \big\rangle_{-\rdt}\Big\} \nonumber \\
	&\propto  \mr{exp}\Big\{\blr{\ln \mathds{1}\big(\rdt\in[\ydt^\low,\ydt^\up]\big)-\notag \\
 &\hspace{4.2cm}\hat{\gamma}_{\mr{d},t}\|\rdt-\mb{H}\xdt\|^2}_{-\rdt}\Big\} \nonumber \\
	&\propto \mathds{1}\big(\rdt\!\in\![\ydt^\low,\ydt^\up]\big)\!\times \mr{exp}\Big\{\!-\hat{\gamma}_{\mr{d},t}\|\mb{r}_{\mr{p},t}\!-\!\lr{\mb{H}}\lr{\xdt}\|^2\Big\}.
	\end{align}
	The update of $\rdt$ is similar to that of $\rpt$. Due to the inherent decoupling of $q(\rdt)$, the variational mean $\lr{\rdmt}$ and variance $\tau_{\rdmt}$ are given by $\ms{F}_r\big(\lr{\mb{H}_{m,:}}\lr{\xdt},\hat{\gamma}_{\mr{d},t}, \ydmt^\low,\ydmt^\up\big)$ and $\ms{G}_r\big(\lr{\mb{H}_{m,:}}\lr{\xdt},\hat{\gamma}_{\mr{d},t}, \ydmt^\low,\ydmt^\up\big)$, respectively.
	
	\textit{3) Updating $\mb{h}_i$.} Taking the expectation of the conditional \eqref{CE-DD-conditional} w.r.t. all latent variables except $\mb{h}_i$, the variational distribution $q(\mb{h}_i)$ is given by 
	\vspace{-0.1cm}
	\setcounter{equation}{33}
	\begin{eqnarray} 
		q(\mb{h}_i) &\propto& \mr{exp}\Big\{\big\langle \ln p(\mb{R}_{\mr{p}}|\mb{H};\hat{\gamma}_{\mr{p}},\mb{X}_{\mr{p}}) + \ln p(\mb{R}_{\mr{d}}|\mb{X}_{\mr{d}},\mb{H};\hat{\bs{\gamma}}_{\mr{d}}) \nonumber \\
		&&\quad\quad\quad + \ln p(\mb{h}_i;\mb{C}_i) \big\rangle_{-\mb{h}_i}\Big\}, 
	\end{eqnarray} 
	which is expanded into \eqref{q-h-i-quantized}. Thus, the variational distribution $q_{\mb{h}_i}(\mb{h}_i)$ is the pdf of a Gaussian random vector with covariance matrix $\bs{\Sigma}_{\mb{h}_i}$ given in \eqref{Sigma-h-noise-quantized} and mean $\lr{\mb{h}_i}$ given in \eqref{mean-h-quantized}. 
	
	\begin{figure*}
	\setcounter{equation}{38}
		\begin{eqnarray}
			q(x_{\mr{d},it})  &\propto& p(x_{\mr{d},it})\, \mr{exp}\bigg\{   -\hat{\gamma}_{\mr{d},t}\bigg[\lr{\|\mb{h}_i\|^2} | x_{\mr{d},it} |^2 - 2\,\Re\bigg\{{\lr{\mb{h}_i^H}}\bigg(\lr{\rdt} - \sum_{j\neq i}^K \lr{\mb{h}_j}\langle x_{\mr{d},jt} \rangle \bigg)x_{\mr{d},it}^*\bigg\}\bigg]  \bigg\} \nonumber \\
			&\propto& p(x_{\mr{d},it})\,\mr{exp}\big\{- \hat{\gamma}_{\mr{d},t}\lr{\|\mb{h}_i\|^2} |x_{\mr{d},it} - z_{\mr{d},it}|^2  \big\} \nonumber \\
			&\propto& p(x_{\mr{d},it})\,\mc{CN}\big(z_{\mr{d},it};\xdit, 1/\big(\hat{\gamma}_{\mr{d},t}\lr{\|\mb{h}_i\|^2}\big)\big).
			\label{eq_qx_est_CSI}
		\end{eqnarray}
			\vspace{-0.2cm}
		\hrulefill
	\end{figure*}
	
	\textit{4) Updating $x_{\mr{d},it}$.} Taking the expectation of the conditional \eqref{CE-DD-conditional} w.r.t. all latent variables except $\xdit$, the variational distribution $q(\xdit)$ is given by 
	\vspace{-0.1cm}
	\setcounter{equation}{37}
	\begin{eqnarray}\label{x-d-i_VB-JED}
	q(\xdit) &\propto& \mr{exp}\Big\{\big\langle \ln p(\mb{r}_{\mr{d},t}|\mb{H},\mb{x}_{\mr{d},t};\hat{\gamma}_{\mr{d},t}) + \ln p(\xdit)\big\rangle_{-\xdit}\!\Big\} \nonumber \\
	&\propto& p(\xdit)\,\mr{exp}\Big\{ \!-\!\hat{\gamma}_{\mr{d},t}\blr{\|\mb{r}_{\mr{d},t}-\mb{H}\mb{x}_{\mr{d},t}\|^2}_{-\xdit}\!\Big\}.
	\end{eqnarray}
	Note that~\eqref{x-d-i_VB-JED} can be expanded into~\eqref{eq_qx_est_CSI}, in which we define
	\setcounter{equation}{39}
	\begin{align} \label{zdit-JED}
	z_{\mr{d},it} &= \frac{\lr{\mb{h}_i^H}}{\lr{\|\mb{h}_i\|^2}}\bigg(\lr{\mb{r}_{\mr{d},t}} - \sum_{j\neq i}^K \lr{\mb{h}_j}\lr{x_{\mr{d},jt}}\bigg) \nonumber \\
	&=\frac{\|\lr{\mb{h}_i}\|^2 \lr{\xdit} + \lr{\mb{h}_i^H}\big(\lr{\mb{r}_{\mr{d},t}} - \lr{\mb{H}}\lr{\mb{x}_{\mr{d},t}}\big)}{\lr{\|\mb{h}_i\|^2}}
	\end{align} 
	as a linear estimate of $x_{\mr{d},it}$. We note that $\lr{\|\mb{h}_i\|^2} = \|\lr{\mb{h}_i}\|^2 + \tr\{\bs{\Sigma}_{\mb{h}_i}\}$. The variational mean and variance of $\xdit$ are given by $\ms{F}_x\big(z_{\mr{d},it}, \hat{\gamma}_{\mr{d},t}\lr{\|\mb{h}_i\|^2}\big)$ and $\ms{G}_x\big(z_{\mr{d},it}, \hat{\gamma}_{\mr{d},t}\lr{\|\mb{h}_i\|^2}\big)$, respectively.
	
	In the \textbf{\textit{M-step}}, the estimates of $\gp$ and $\gdt$ are updated to maximize $\ln p(\Yp,\Yd,\Rp,\Rd,\mb{X}_{\mr{d}},\mb{H};{\gamma}_{\mr{p}},\bs{\gamma}_{\mr{d}}, \mb{X}_{\mr{p}},\{\mb{C}_i\})$ w.r.t. the variational distribution $q(\Rp,\Rd,\Xd,\mb{H})$, i.e.,
	\vspace{-0.05cm}
	\begin{align}
	    \hat{\gamma}_{\mr{p}} &= \arg\max_{\gp}\; \blr{\ln p(\Rp|\mb{H};\gp,\Xp)} \nonumber \\
	    &= \arg\max_{\gp}\; M\Tp - \gp \blr{\|\mb{R}_{\mr{p}}-\mb{H}\mb{X}_{\mr{p}}\|^2} \nonumber \\
	    & = 
	   \frac{MT_{\mr{p}}} {\sum_{t=1}^{\Tp}\blr{\|\rpt-\mb{H}\xpt\|^2}}\label{gamma-p}
	\end{align}
	and for $t=1,\ldots,\Td$
	\vspace{-0.05cm}
	\begin{align}
	    \hat{\gamma}_{\mr{d,t}} &= \arg\max_{\gdt}\; \blr{\ln p(\rdt|\mb{H},\xdt;\gdt} \nonumber \\
	    &= \arg\max_{\gdt}\; \Tp - \gdt \blr{\|\rdt-\mb{H}\xdt\|^2} \nonumber \\
	    & = \frac{T_{\mr{p}}} {\blr{\|\rdt-\mb{H}\xdt\|^2}}.\label{gamma-d}
	\end{align}
	
   By applying Theorem \ref{theorem-1}, we have
   	\vspace{-0.05cm}
    \begin{align}\label{gamma-p-exp}
        \blr{\|\mb{r}_{\mr{p},t}-\mb{H}\mb{x}_{\mr{p},t}\|^2} &= \|\lr{\mb{r}_{\mr{p},t}}-\lr{\mb{H}}{\mb{x}_{\mr{p},t}}\|^2 + \sum_{m=1}^M \tau_{\rpmt} \nonumber \\
        &\quad+ \sum_{i=1}^K {|x_{\mr{p},it}|^2}\tr\{\bs{\Sigma}_{\mb{h}_i}\}
    \end{align}
and      
\vspace{-0.15cm}
\begin{align}\label{gamma-dt-exp}
	\blr{\|\mb{r}_{\mr{d},t}-\mb{H}\mb{x}_{\mr{d},t}\|^2} &= \|\lr{\mb{r}_{\mr{d},t}}-\lr{\mb{H}}\lr{\mb{x}_{\mr{d},t}}\|^2 + \sum_{m=1}^M \tau_{\rdmt} \nonumber \\
	&+\sum_{i=1}^K \left[\lr{|x_{\mr{d},it}|^2}\tr\{\bs{\Sigma}_{\mb{h}_i}\} + \tau_{x_{i}}\|\lr{\mb{h}_i}\|^2\right].
	\end{align}
	
    By iteratively optimizing $\{q(\rpt)\}$, $\{q(\rdt)\}$, $\{q(\mb{h}_i)\}$, $\{q(\xdit)\}$, $\hat{\gamma}_{\mr{p}}$, and $\{\hat{\gamma}_{\mr{d},t}\}$, we obtain the VBEM algorithm for estimating $\Rp$, $\Rd$, $\mb{H}$, $\Xd$, $\gp$, and $\{\gdt\}$. We refer to this scheme as the \textit{\textbf{MF-QVB-JED algorithm}} for joint channel estimation and data detection. 

    \textit{\textbf{Remark 1:} If $\gp$ and $\{\gdt\}$ are all set to $N_0^{-1}$, the MF-QVB-JED algorithm is equivalent to the VB-based joint channel estimation and data detection approach in \cite{Thoota2021Variational}. We will refer to the scheme in \cite{Thoota2021Variational} as the \textbf{conv-QVB-JED algorithm}. In conv-QVB-JED, the variational covariance matrix of $\mb{h}_i$ given by
        \vspace{-0.1cm}
    \begin{align}\label{Sigma-h-i-VB}
	\bs{\Sigma}_{\mb{h}_i} = \bigg[N_0^{-1}\bigg(\sum_{t=1}^{\Tp} |\xpit|^2 + \sum_{t=1}^{T_{\mr{d}}} \lr{|x_{\mr{d},it}|^2}\bigg)\mb{I}_M + \mb{C}_i^{-1}\bigg]^{-1}
    \end{align}
    becomes smaller with increasing $\Td$ or $\lr{|x_{\mr{d},it}|^2}, \forall t$. This result, however, implies that the estimation of $\mb{h}_i$ becomes more accurate with a longer transmission phase or even with an unreliable estimate of $\xdit$ reflected through large $\tau_{\xdit}$ (and $\lr{|x_{\mr{d},it}|^2}$). In MF-QVB-JED, an unreliable estimation of $\xdit$ will decrease $\hat{\gamma}_{\mr{d,t}}$ in \eqref{gamma-d}. Evidently, the effect of $\lr{|x_{\mr{d},it}|^2}$ on the variational covariance matrix $\bs{\Sigma}_{\mb{h}_i}$ of $\mb{h}_i$ in \eqref{Sigma-h-noise-quantized} is less important than its effect on $\bs{\Sigma}_{\mb{h}_i}$ in \eqref{Sigma-h-i-VB}. Therefore, in the MF-QVB-JED algorithm, an unreliable estimate of $\xdit$ will not increase the accuracy of estimating $\mb{h}_i$. This is one of explanations for the superior performance of MF-QVB-JED compared with conv-QVB-JED.}

    \textit{\textbf{Remark 2:} By denoting 
    \vspace{-0.1cm}
    \begin{align}
        \gamma_i &= \hat{\gamma}_{\mr{p}} \sum_{t=1}^{\Tp} |\xpit|^2 + \sum_{t=1}^{T_{\mr{d}}} \hat{\gamma}_{\mr{d},t}\lr{|x_{\mr{d},it}|^2}
    \end{align}
    and 
    \vspace{-0.15cm}
    \begin{align} 
        \mb{k}_i &= \gamma_i^{-1} \Bigg[\hat{\gamma}_{\mr{p}}\sum_{t=1}^{\Tp} \!\bigg(\!\lr{\mb{r}_{\mr{p},t}}\! -\! \sum_{j\neq i}^K \langle\mb{h}_j\rangle \xpjt\! \bigg) \xpit^*  \nonumber \\
        &\quad\quad +  \sum_{t=1}^{T_{\mr{d}}} \hat{\gamma}_{\mr{d},t} \bigg(\lr{\mb{r}_{\mr{d},t}} - \!\sum_{j\neq i}^K \langle\mb{h}_j\rangle\langle x_{\mr{d},jt} \rangle \bigg)\lr{x_{\mr{d},it}^*}\Bigg],
    \end{align}
    we note that the variational distribution $q(\mb{h}_i)$ in \eqref{q-h-i-quantized} can also be expressed as 
    \vspace{-0.1cm}
    \begin{align*}
    q(\mb{h}_i) &\propto \mc{CN}(\mb{h}_i;\mb{k}_i,\gamma_i^{-1}\mb{I}_M)\,\mc{CN}(\mb{h}_i;\mb{0},\mb{C}_i) \nonumber  \\
    &= \mc{CN}\big(\mb{h}_i;\gamma_i\big(\gamma_i\mb{I}_M + \mb{C}_i^{-1}\big)^{-1}\mb{k}_i,\big(\gamma_i\mb{I}_M + \mb{C}_i^{-1}\big)^{-1}\big), 
    \end{align*}
    which then explains the results $\bs{\Sigma}_{\mb{h}_i} = \big(\gamma_i\mb{I}_M + \mb{C}_i^{-1}\big)^{-1}$ and $\lr{\mb{h}_i} = \gamma_i\bs{\Sigma}_{\mb{h}_i}\mb{k}_i$ in \eqref{Sigma-h-noise-quantized} and \eqref{mean-h-quantized}, respectively. We also note that $\mb{k}_i$ can be written as
    \begin{align*}
        \mb{k}_i
        =&\bigg(1-\frac{\sum_{t=1}^{\Td} \hat{\gamma}_{\mr{d},t}\tau_{\xdit}}{\gamma_i}\bigg)\lr{\mb{h}_i} \nonumber \\
        &+ \gamma_i^{-1}\Big[\hat{\gamma}_{\mr{p}} \mb{E}_{\mr{p}}[\Xp]_{i,:}^H + \mb{E}_{\mr{d}}\big([\lr{\Xp}]_{i,:}^H\odot\hat{\bs{\gamma}}_{\mr{d}}\big) \Big],
    \end{align*}
    enabling its efficient computation using the residual matrices $\mb{E}_{\mr{p}} = \lr{\Rp} - \lr{\mb{H}}\Xp$ and $\mb{E}_{\mr{d}} = \lr{\Rd} - \lr{\mb{H}}\lr{\Xd}$.
    } 

	\begin{algorithm}[t]
	\caption{-- MF-QVB-JED for Few-Bit MIMO Joint Channel Estimation and Data Detection}
	\begin{algorithmic}[1]
		\State \textbf{Input:} $\mb{Y}_{\mr{p}}, \mb{Y}_{\mr{d}}, \mb{X}_{\mr{p}}, \mb{C}_{i},\forall i$
		\State \textbf{Output:} $\hat{\mb{H}}, \hat{\mb{X}}_{\mr{d}}$
		\State Initialize $\hat{\mb{H}}^1 = \mb{0}$, $\hat{\mb{X}}_{\mr{d}}^1 = \mb{0}$, $\tau_{\xdit}^1 = \mr{Var}_{p(x_i)}[x_i], \forall i,\forall t$, $\hat{\mb{R}}_{\mr{p}}^1 = \mb{Y}_{\mr{p}}$, $\hat{\mb{R}}_{\mr{d}}^1 = \mb{Y}_{\mr{d}}$,  $\tau_{\rpmt}^1 = 0$, $\tau_{\rdmt}^1 = 0,\forall m, \forall t$, $\mb{E}_{\mr{p}} = \hat{\mb{R}}_{\mr{p}}^1 - \hat{\mb{H}}^1\Xp$, and $\mb{E}_{\mr{d}} = \hat{\mb{R}}_{\mr{d}}^1 - \hat{\mb{H}}^1\hat{\mb{X}}_{\mr{d}}^1$\;
		\For{$\ell=1,2,\ldots$}
		    \State Update $\hat{\gamma}_{\mr{p}}^{\ell}$ using \eqref{gamma-p} and \eqref{gamma-p-exp}
		    \For{$t=1,\ldots,\Tp$}\Comment{Update $\Rp$}
				\State $\mb{s}_{\mr{p},t}^\ell \gets \hat{\mb{r}}_{\mr{p},t}^\ell - [\mb{E}_{\mr{p}}]_{:,t}$
				\State $\hphantom{\mb{s}_{\mr{p},t}^\ell}\mathllap{\hat{\mb{r}}_{\mr{p},t}^{\ell+1}} \gets \ms{F}_r\big(\mb{s}_{\mr{p},t}^\ell ,\hat{\gamma}_{\mr{p}}^{\ell},\mb{y}_{\mr{p},t}^{\low}, \mb{y}_{\mr{p},t}^\up\big)$
		        \State $\hphantom{\mb{s}_{\mr{p},t}^\ell}\mathllap{\bs{\tau}_{\mb{r}_{\mr{p},t}}^{\ell+1}} \gets \ms{G}_r\big(\mb{s}_{\mr{p},t}^\ell ,\hat{\gamma}_{\mr{p}}^{\ell},\mb{y}_{\mr{p},t}^{\low}, \mb{y}_{\mr{p},t}^\up\big)$
	       \EndFor
	       \State $\mb{E}_{\mr{p}} \gets \mb{E}_{\mr{p}} - \hat{\mb{R}}_{\mr{p}}^\ell + \hat{\mb{R}}_{\mr{p}}^{\ell+1}$
	        \For{$t=1,\ldots,\Td$}\Comment{Update $\Rd$}
	        \State Update $\hat{\gamma}_{\mr{d},t}^{\ell}$ using \eqref{gamma-d} and \eqref{gamma-dt-exp}
		    \State $\mb{s}_{\mr{d},t}^\ell \gets \hat{\mb{r}}_{\mr{d},t}^\ell - [\mb{E}_{\mr{d}}]_{t,:}$
		    \State $\hphantom{\mb{s}_{\mr{d},t}^\ell}\mathllap{\hat{\mb{r}}_{\mr{d},t}^{\ell+1}} \gets \ms{F}_r\big(\mb{s}_{\mr{d},t}^\ell, \hat{\gamma}_{\mr{d},t}^{\ell},\mb{y}_{\mr{d},t}^{\low}, \mb{y}_{\mr{d},t}^\up\big)$
		    \State $\hphantom{\mb{s}_{\mr{d},t}^\ell}\mathllap{\bs{\tau}_{\mb{r}_{\mr{d},t}}^{\ell+1}} \gets \ms{G}_r\big(\mb{s}_{\mr{d},t}^\ell, \hat{\gamma}_{\mr{d},t}^{\ell},\mb{y}_{\mr{d},t}^{\low}, \mb{y}_{\mr{d},t}^\up\big)$
		    \EndFor 
		    \State $\mb{E}_{\mr{d}} \gets \mb{E}_{\mr{d}} - \hat{\mb{R}}_{\mr{d}}^\ell + \hat{\mb{R}}_{\mr{d}}^{\ell+1}$
			\For{$i=1,\ldots,K$}\Comment{Update $\mb{H}$}
		    \State $\gamma_i^{\ell} \gets \hat{\gamma}_{\mr{p}}^{\ell} \sum_{t=1}^{\Tp} |\xpit|^2 + \sum_{t=1}^{\Td} \hat{\gamma}_{\mr{d},t}^{\ell} \lr{|x_{\mr{d},it}^{\ell}|^2}$
		    \State $\hphantom{\gamma_i^{\ell}}\mathllap{\mb{k}^{\ell}_i} \gets \bigg(1-\frac{\sum_{t=1}^{\Td} \hat{\gamma}_{\mr{d},t}^{\ell} \tau_{\xdit}^\ell}{\gamma_i^\ell}\bigg)\hat{\mb{h}}_i^{\ell}$
		    \NoNumber{\quad\quad\quad $+\, (\gamma_i^{\ell})^{-1}\Big[{\hat{\gamma}_{\mr{p}}^{\ell}} \mb{E}_{\mr{p}}[\Xp]_{i,:}^H + \mb{E}_{\mr{d}}\big([\Xp^\ell]_{i,:}^H\odot\hat{\bs{\gamma}}_{\mr{d}}^\ell\big) \Big]$}
		    \State $\hphantom{\gamma_i^{\ell}}\mathllap{\bs{\Sigma}_{\mb{h}_i}^{\ell+1}} \gets \big(\gamma_i^{\ell}\mb{I}_M + \mb{C}_i^{-1}\big)^{-1}$
		    \State $\hphantom{\gamma_i^{\ell}}\mathllap{\hat{\mb{h}}_i^{\ell+1}} \gets \gamma_i^\ell\bs{\Sigma}_{\mb{h}_i}^{\ell+1}\mb{k}_i^\ell$
			\State $\hphantom{\gamma_i^{\ell}}\mathllap{\mb{E}_{\mr{p}}} \gets\mb{E}_{\mr{p}} + (\hat{\mb{h}}_i^\ell - \hat{\mb{h}}_i^{\ell+1})[\Xp]_{i,:}$
			\State $\hphantom{\gamma_i^{\ell}}\mathllap{\mb{E}_{\mr{d}}} \gets \mb{E}_{\mr{d}} + (\hat{\mb{h}}_i^\ell - \hat{\mb{h}}_i^{\ell+1})[\hat{\mb{X}}_{\mr{d}}^\ell]_{i,:}$\;
			\EndFor
			\For{$t=1,\ldots,\Td$} \Comment{Update $\Xd$} 
			\For{$i=1,\ldots,K$}
			\State $z_{it}^\ell \gets \frac{\|\hat{\mb{h}}_i^{\ell}\|^2\hat{x}_{\mr{d},it}^\ell + (\hat{\mb{h}}_i^{\ell})^H[\mb{E}_{\mr{d}}]_{:,t}}{\|\hat{\mb{h}}_i^{\ell}\|^2 + \tr\{\bs{\Sigma}_{\mb{h}_i}^{\ell}\}}$
			\State $\hphantom{z_{it}^\ell}\mathllap{\hat{x}_{\mr{d},it}^{\ell+1}} \gets \ms{F}_x\big(z_{it}^\ell, \hat{\gamma}_{\mr{d},t}^{\ell}\big(\|\hat{\mb{h}}_i^{\ell}\|^2 +\tr\{\bs{\Sigma}_{\mb{h}_i}^{\ell}\}\big)\big)$
			\State $\hphantom{z_{it}^\ell}\mathllap{\tau_{\xdit}^{\ell+1}} \gets \ms{G}_x\big(z_{it}^\ell, \hat{\gamma}_{\mr{d},t}^{\ell}\big(\|\hat{\mb{h}}_i^{\ell}\|^2 +\tr\{\bs{\Sigma}_{\hat{\mb{h}}_i}^{\ell}\}\big) \big)$
			\State $\hphantom{z_{it}^\ell}\mathllap{[\mb{E}_{\mr{d}}]_{:,t}} \gets [\mb{E}_{\mr{d}}]_{:,t} + \hat{\mb{h}}_i^{\ell}(\hat{x}_{\mr{d},it}^\ell - \hat{x}_{\mr{d},it}^{\ell+1})$
			\EndFor
			\EndFor
		\EndFor
		\State $\forall t, \forall i: \hat{x}_{\mr{d},it} \!\gets\! \argmax_{a\in\mc{S}} p_a\mc{CN}\big(z_{it}^\ell;a, 1/[\hat{\gamma}_{\mr{d},t}^{\ell}(\|\hat{\mb{h}}_i^{\ell}\|^2 + \tr\{\bs{\Sigma}_{\mb{h}_i}^{\ell}\})]\big)$.
		\end{algorithmic}
		\label{algo-3}
	\end{algorithm}
	
 The proposed MF-QVB-JED algorithm is summarized Algorithm \ref{algo-3}. Here, we use $\hat{\mb{r}}_{\mr{p},t}^{\ell}$,  $\hat{\mb{r}}_{\mr{d},t}^{\ell}$, $\hat{\mb{h}}_i^{\ell}$, $\hat{x}_i^{\ell}$, $\hat{\gamma}_{\mr{p}}^{\ell}$, and $\hat{\gamma}_{\mr{d},t}^{\ell}$ to replace $\lr{\rpt}$, $\lr{\rdt}$, $\lr{\mb{h}_i}$, $\lr{x_i}$, $\hat{\gamma}_{\mr{p}}$, and $\hat{\gamma}_{\mr{d},t}$ at iteration $\ell$. We also include in the algorithm the residual terms $\mb{E}_{\mr{p}}$ and $\mb{E}_{\mr{d}}$, which are adjusted to reflect any update to the estimates of $\Rp$, $\Rd$, $\mb{H}$, and $\Xd$. 
    \begin{figure*}
	\begin{align}\label{LMMSE-CE-DD-conditional}
		&p(\Yp,\Yd,\Rp,\Rd,\mb{X}_{\mr{d}},\mb{H};{\gamma}_{\mr{p}},\{\bs{\Gamma}_{t}\},\mb{X}_{\mr{p}},\{\mb{C}_i\}) \nonumber \\
		&= p(\Yp|\Rp)p(\Rp|\mb{H}; \gp,\mb{X}_{\mr{p}}) p(\Yd|\Rd)p(\Rd|\mb{X}_{\mr{d}},\mb{H};\{\bs{\Gamma}_{t}\}) p(\mb{X}_{\mr{d}}) 
		p(\mb{H}|\{\mb{C}_i\}) \nonumber \\
		& = \Bigg[\prod_{t=1}^{\Tp} p(\ypt|\rpt)p(\rpt|\mb{H};\gp,\xpt) \Bigg] \left[\prod_{t=1}^{\Td} p(\ydt|\rdt)p(\rdt|\mb{H},\xdt;\bs{\Gamma}_{t}) p(\xdt)\right]  \prod_{i=1}^Kp(\mb{h}_i|\mb{C}_i)
	\end{align}
	\hrule
\end{figure*}

	\subsection{Proposed LMMSE-QVB-JED Algorithm} \label{sec-LMMSE-QVB-JED}
	This section extends the LMMSE-QVB algorithm to the case of joint channel estimation and data detection. We denote $\gp$ and $\{\bs{\Gamma}_t\} = \{\bs{\Gamma}_1,\ldots,\bs{\Gamma}_{\Td}\}$ as the precision of the noise during the pilot transmission phase and the $\Td$ data transmission time slots, respectively. The joint distribution of all the observations and latent variables in \eqref{CE-DD-conditional} are now factored as given in \eqref{LMMSE-CE-DD-conditional}, where $p(\rpt|\mb{H};\gp,\xpt) = \mc{CN}\big(\rpt;\mb{H}\xpt,\gp^{-1}\mb{I}_M\big)$ and $p(\rdt|\mb{H},\mb{x}_{\mr{d},t};\bs{\Gamma}_t) = \mc{CN}\big(\rdt;\mb{H}\xdt,\bs{\Gamma}_t^{-1}\big)$.

    In the \textbf{\textit{E-step}}, for currently fixed estimates $\hat{\gamma}_{\mr{p}}$ and $\{\hat{\bs{\Gamma}}_t\}$ of $\gp$ and $\{\bs{\Gamma}_t\}$, respectively, we aim to obtain the mean field variational distribution $q(\Rp,\Rd,\mb{H},\Xd)$ of $\Rp$, $\Rd$,  $\mb{H}$, and $\Xd$ given $\Yp$ and $\Yd$ such that
	\begin{align}
		&p\big(\Rp,\Rd,\mb{X}_{\mr{d}},\mb{H}|\Yp,\Yd;\hat{\gamma}_{\mr{p}},\{\hat{\bs{\Gamma}}_{t}\},\Xp,\{\mb{C}_i\}\big) \nonumber \\
		&\approx q\big(\Rp,\Rd,\mb{X}_{\mr{d}},\mb{H}\big) \nonumber \\
		&= \Bigg[\prod_{t=1}^{T_{\mr{p}}}q(\rpt)\Bigg]\!\left[ \prod_{t=1}^{T_{\mr{d}}} q(\rdt)\right]\! \left[\prod_{i=1}^Kq(\mb{h}_i)\right]\! \left[\prod_{i=1}^K\prod_{t=1}^{T_{\mr{d}}} q(x_{\mr{d},it})\right].
	\end{align}
	
	\textit{1) Updating $\rpt$.} Similar to the MF-QVB-JED algorithm. 
	
	\textit{2) Updating $\rdmt$.} Similar to the LMMSE-QVB algorithm, the variational mean $\lr{\rdmt}$ and variance $\tau_{\rdmt}$ are determined by $\ms{F}_r\big(s_{\mr{d},mt},\hat{\gamma}_{t,mm}, \ydmt^\low,\ydmt^\up\big)$ and $\ms{G}_r\big(s_{\mr{d},mt},\hat{\gamma}_{t,mm}, \ydmt^\low,\ydmt^\up\big)$,
	where
	\begin{align}
	    s_{\mr{d},mt} = \lr{\rdmt} - \frac{\hat{\bs{\Gamma}}_{t,m}^H}{\hat{\gamma}_{t,mm}}\big(\lr{\rdt} - \lr{\mb{H}}\lr{\xdt}\big)
	\end{align}
	and $\hat{\bs{\Gamma}}_{t,m}$ and $\hat{\gamma}_{t,mm}$ are the $m$th column and the $(m,m)$-element of $\hat{\bs{\Gamma}}_t$, respectively. 
	
	\textit{3) Updating $\mb{h}_i$.} Taking the expectation of the conditional \eqref{LMMSE-CE-DD-conditional} w.r.t. all latent variables except $\mb{h}_i$, the variational distribution $q(\mb{h}_i)$ is given by 
	\begin{align} 
		q(\mb{h}_i) &\propto \mr{exp}\Big\{\!\blr{\ln p(\mb{R}_{\mr{p}}|\mb{H};\hat{\gamma}_{\mr{p}},\mb{X}_{\mr{p}}) + \ln p(\mb{R}_{\mr{d}}|\mb{X}_{\mr{d}},\mb{H};\{\hat{\bs{\Gamma}}_{t}\})  \nonumber \\
		&\quad\quad\quad + \ln p(\mb{h}_i;\mb{C}_i)}_{-\mb{h}_i}\Big\}.
	\end{align} 
	Following the same procedure to obtain $q(\mb{h}_i)$ as in the MF-QVB-JED algorithm, we have 
	\begin{align} 
		q(\mb{h}_i) &\propto \mc{CN}\big(\mb{h}_i;\mb{k}_i,\bs{\Gamma}^{-1}\big)\mc{CN}(\mb{h}_i;\mb{0},\mb{C}_i) \nonumber \\
		&= \mc{CN}\big(\mb{h}_i;\big(\bs{\Gamma}_i + \mb{C}_i^{-1}\big)^{-1}\bs{\Gamma}_i\mb{k}_i, \big(\bs{\Gamma}_i + \mb{C}_i^{-1}\big)^{-1}\big),
	\end{align} 
	where $\bs{\Gamma}_i$ and $\mb{k}_i$ are defined as
	\begin{align}\label{Gamma-i}
	\bs{\Gamma}_i &= \hat{\gamma}_{\mr{p}}\sum_{t=1}^{\Tp} |\xpit|^2\mb{I}_M + \sum_{t=1}^{T_{\mr{d}}} \lr{|x_{\mr{d},it}|^2} \hat{\bs{\Gamma}}_t\\
	    \mb{k}_i &= \bigg(\mb{I}_M - \bs{\Gamma}_i^{-1}\sum_{t=1}^{\Td}\hat{\bs{\Gamma}}_t\tau_{\xdit}\bigg)\lr{\mb{h}_i} \nonumber \\
	    &\quad+ \bs{\Gamma}_i^{-1}\bigg[\hat{\gamma}_{\mr{p}} \mb{E}_{\mr{p}}\mb{x}_{\mr{p},i}^* + \sum_{t=1}^{\Td} \hat{\bs{\Gamma}}_t \mb{e}_{\mr{d},t}\xdit^* \bigg]
	\end{align}
	and where $\mb{E}_{\mr{p}} = \lr{\Rp} - \lr{\mb{H}}\Xp$ and $\mb{e}_{\mr{d},t} = \lr{\rdt} - \lr{\mb{H}}\lr{\xdt}$ are the residual terms. The variational covariance matrix and mean of $\mb{h}_i$ are now given by $\bs{\Sigma}_{\mb{h}_i} = \big(\bs{\Gamma}_i + \mb{C}_i^{-1}\big)^{-1}$ and  $\lr{\mb{h}_i} = \bs{\Sigma}_{\mb{h}_i}\bs{\Gamma}_i\mb{k}_i$.                                                                                                                                                                                                                              
	
	\textit{4) Updating $x_{\mr{d},it}$.} Taking the expectation of the conditional \eqref{CE-DD-conditional} w.r.t. all latent variables except $\xdit$, the variational distribution $q(\xdit)$ is given by  
	\begin{align}\label{x-d-i_LMMSE-QVB-JED}
	q(\xdit) \propto \mr{exp}\Big\{\!\big\langle \!\ln p(\mb{r}_{\mr{d},t}|\mb{H},\xdt;\hat{\bs{\Gamma}}_{t})\! + \ln p(\xdit)\big\rangle_{-\xdit}\!\Big\}. 
	\end{align}
	Similar to the procedure in the LMMSE-QVB and MF-QVB-JED algorithms, we obtain
	\begin{align}
	q(\xdit) \propto p(\xdit)\,\mc{CN}\big(z_{\mr{d},it};\xdit,1/\lr{\mb{h}_i^H\hat{\bs{\Gamma}}_t\mb{h}_i}\big),
	\end{align}
    where
	\begin{align} 
	z_{\mr{d},it} &=  \frac{\lr{\mb{h}_i^H}\hat{\bs{\Gamma}}_t}{\lr{\mb{h}^H_i\hat{\bs{\Gamma}}_t\mb{h}_i}}\bigg(\lr{\mb{r}_{\mr{d},t}} - \sum_{j\neq i}^K \lr{\mb{h}_j}\lr{x_{\mr{d},jt}}\bigg) \nonumber \\
	&= \frac{\lr{\mb{h}_i^H}\hat{\bs{\Gamma}}_t\lr{\mb{h}_i}\lr{\xdit} + \lr{\mb{h}_i^H}\hat{\bs{\Gamma}}_t\mb{e}_{\mr{d},t}}{\lr{\mb{h}^H_i\hat{\bs{\Gamma}}_t\mb{h}_i}}
	\end{align} 
	is a linear estimate of $x_{\mr{d},it}$. We note that $\lr{\mb{h}^H_i\hat{\bs{\Gamma}}_t\mb{h}_i}  = \lr{\mb{h}_i}\hat{\bs{\Gamma}}_t \lr{\mb{h}_i} + \tr\{\hat{\bs{\Gamma}}_t\bs{\Sigma}_{\mb{h}_i}\}$.
	
	In the \textbf{\textit{M-step}}, $\gp$ and $\{\bs{\Gamma}_t\}$ are estimated to maximize $\ln p(\Yp,\Yd,\Rp,\Rd,\mb{X}_{\mr{d}},\mb{H};{\gamma}_{\mr{p}},\{\bs{\Gamma}_t\}, \mb{X}_{\mr{p}},\{\mb{C}_i\})$ w.r.t. the variational distribution $q(\Rp,\Rd,\Xd,\mb{H})$. The update of $\hat{\gamma}_{\mr{p}}$ is similar to the procedure in the MF-QVB-JED algorithm and is given in \eqref{gamma-p} and \eqref{gamma-p-exp}. The update of $\hat{\bs{\Gamma}}_t$ is given by  
	\begin{align*}
	\hat{\bs{\Gamma}}_t &= \arg\max_{\bs{\Gamma}_t}\; \blr{\ln p(\rdt|\mb{H},\xdt;\bs{\Gamma}_t)} \nonumber \\
	&=\arg\max_{\bs{\Gamma}_t}\; \Big[\ln|\bs{\Gamma}_t| \nonumber \\
	&\quad\quad\quad\quad\quad\;\; - \blr{(\rdt-\mb{H}\xdt)^H\bs{\Gamma}_t(\rdt-\mb{H}\xdt)}\Big].
	\end{align*}
	Applying Theorem \ref{theorem-1}, we have 
	\begin{align} 
	&\blr{(\rdt-\mb{H}\xdt)^H\bs{\Gamma}_t(\rdt-\mb{H}\xdt)} \nonumber \\
	&=\tr\Big\{\Big[\big(\lr{\rdt}-\lr{\mb{H}}\lr{\xdt}\big)\big(\lr{\rdt}-\lr{\mb{H}}\lr{\xdt}\big)^H+\bs{\Sigma}_{\rdt} \nonumber \\
	&\quad\quad\quad +\sum_{i=1}^K \lr{|\xdit|^2}\bs{\Sigma}_{\mb{h}_i} + \lr{\mb{H}}\bs{\Sigma}_{\xdt}\lr{\mb{H}}^H\Big]\bs{\Gamma}_t\Big\} .
	\end{align}
	Due to the rank deficiency of $\blr{(\rdt-\mb{H}\xdt)^H\bs{\Gamma}_t(\rdt-\mb{H}\xdt)}_{-\bs{\Gamma}_t}$, we propose to use the following estimator for $\bs{\Gamma}_t$:
    \begin{align}\label{W-t}
	\hat{\bs{\Gamma}}_t = \bigg(&\frac{\|\lr{\rdt} - \lr{\mb{H}}\lr{\xdt}\|^2}{M}\mb{I}_M +\bs{\Sigma}_{\rdt}  \nonumber \\
	&+\sum_{i=1}^K \lr{|\xdit|^2}\bs{\Sigma}_{\mb{h}_i} + \lr{\mb{H}}\bs{\Sigma}_{\xdt}\lr{\mb{H}}^H\bigg)^{-1}.
	\end{align}
	
By iteratively optimizing $\{q(\rpt)\}$, $\{q(\rdt)\}$, $\{q(\mb{h}_i)\}$, $\{q(\xdit)\}$, $\hat{\gamma}_{\mr{p}}$, and $\{\hat{\bs{\Gamma}}_{t}\}$, we obtain the VBEM algorithm for estimating $\Rp$, $\Rd$, $\mb{H}$, $\Xd$, $\gp$, and $\{\bs{\Gamma}_t\}$. We refer to this scheme as the \textit{\textbf{LMMSE-QVB-JED algorithm}} for joint channel estimation and data detection. The implementation of the LMMSE-QVB-JED algorithm is similar to that of MF-QVB-JED presented in Algorithm \ref{algo-3}. We skip the summary of the LMMSE-QVB-JED algorithm for brevity.
	
	\subsection{Practical Aspects of Implementing MF/LMMSE-QVB-JED}
    
    \textit{1) Computing $\lr{|\xdit|^2}$:} For PSK signaling, the variational second moment $\lr{|\xdit|^2}$ is constant and need not be updated in each iteration of the algorithms. We present the proof for this observation in Appendix \ref{append-B}.
    
    \textit{2) Computing $\bs{\Sigma}_{\mb{h}_i}$ with uncorrelated channels:} When $\mb{C}_i$ is a diagonal matrix, the variational covariance matrix $\bs{\Sigma}_{\mb{h}_i}$ in \eqref{Sigma-h-i-VB} is also a diagonal matrix and its computation does not require matrix inversion. Thus, the MF-QVB-JED algorithm can be implemented without any matrix inversion. This property does not hold for the LMMSE-QVB-JED algorithm, since $\bs{\Gamma}_i$ in \eqref{Gamma-i} is not in general a diagonal matrix.
    
    \textit{3) Lite implementation of MF-QVB-JED:} Instead of using the latent variable $\gdt$ as the precision at time slot $t$, we can impose a single latent variable $\gamma_{\mr{d}}$ as the precision for all time slots. A lite version of MF-QVB-JED can be devised using the same procedure as in Section \ref{sec-MF-QVB-JED} where $\{\gdt\}$ is replaced by $\gamma_{\mr{d}}$. In the \textbf{\textit{M-step}}, the estimate of $\gamma_{\mr{d}}$ can be found as 
    \begin{align}
        \hat{\gamma}_{\mr{d}} = \frac{M\Td}{\sum_{t=1}^{\Td} \lr{\|\rdt-\mb{H}\xdt\|^2}},
    \end{align}
    where $\lr{\|\rdt-\mb{H}\xdt\|^2}$ is given in \eqref{gamma-dt-exp}.
    
    \textit{5) Lite implementation of LMMSE-QVB-JED:} Instead of using the latent variable $\bs{\Gamma}_t$ as the precision matrix at the time slot $t$, we could use the same precision matrix $\bs{\Gamma}$ for all data time slots. A lite version of LMMSE-QVB-JED can be devised using the same procedure as in Section \ref{sec-LMMSE-QVB-JED} where $\{\bs{\Gamma}_t\}$ is replaced by $\bs{\Gamma}$. In the  \textbf{\textit{M-step}}, we propose to use the following estimator for $\bs{\Gamma}$:
	\begin{align}\label{W-joint-2}
	\hat{\bs{\Gamma}} =\Td&\bigg(\frac{\|\lr{\Rd}- \lr{\mb{H}}\lr{\Xd}\|_F^2}{M}\mb{I}_M +\bs{\Sigma}_{\mb{r}_{\mr{d}}}  \nonumber \\
	&+\sum_{i=1}^K \lr{\|\mb{x}_{\mr{d},i}\|^2}\bs{\Sigma}_{\mb{h}_i} +  \lr{\mb{H}}\bs{\Sigma}_{\mb{x}_{\mr{d}}}\lr{\mb{H}}^H\bigg)^{-1} ,
	\end{align}
	where we denote $\bs{\Sigma}_{\mb{r}_{\mr{d}}} = \sum_{t=1}^{\Td} \bs{\Sigma}_{\rdt}$, $\bs{\Sigma}_{\mb{x}_{\mr{d}}} = \sum_{t=1}^{\Td} \bs{\Sigma}_{\xdt}$ and $\lr{\|\mb{x}_{\mr{d},i}\|^2} = \sum_{t=1}^{\Td} \lr{|\xdit|^2}$.
	
	We observe in our simulations that the lite version of MF/LMMSE-QVB-JED slightly increases the detection error compared to the original version. However, the lite version can significantly reduce the computational complexity, especially for the LMMSE-QVB-JED algorithm. LMMSE-QVB-JED requires one matrix inversion in \eqref{W-joint-2} for computing $\hat{\bs{\Gamma}}$ in the lite version, while requiring $\Td$ matrix inversions to compute $\{\hat{\bs{\Gamma}}_t\}$ in the original version. In the numerical results, we will use the lite version of these algorithms.
	
	\section{Numerical Results}
	\label{sec:simulation_results}
	This section presents numerical results comparing the performance of the proposed VB-based methods with the conventional quantized VB-based method, denoted as conv-QVB, in~\cite{Thoota2021Variational} and FBM-DetNet in~\cite{Ly2022TWC}, which are the most recent and related methods to the work in this paper. The maximum number of iterations is set to $50$ for all the iterative algorithms. The covariance matrices $\mb{C}_i$ are normalized such that their diagonal elements are 1, which implies $\mathbb{E}[\|\mb{h}_i\|^2] = M, \; \forall i$. The noise variance $N_0$ is set according to the operating SNR, which is defined as
    \begin{equation}
        \mr{SNR} = \frac{\mathbb{E}[\|\mathbf{Hx}\|^2]}{\mathbb{E}[\|\mb{n}\|^2]} = \frac{\sum_{i=1}^{K}\tr\{\mb{C}_i\}}{MN_0} = \frac{K}{N_0}.
    \end{equation}
    For i.i.d. channels, we set $\mb{C}_i = \mb{I}, \; \forall i$. For spatially correlated channels, we use the typical urban channel model in~\cite{li2017channel} where the power angle spectrum of the channel model follows a Laplacian distribution with an angle spread of $10^{\circ}$. The covariance matrix $\mb{C}_i$ is obtained according to~\cite[Eq. (2)]{Li2015Pilot}. Unless otherwise stated, we set the training length $\Tp = 2K$ and the data transmission length $\Td = 100$.

\begin{figure}[t!]
    \centering
    \includegraphics[width=\linewidth]{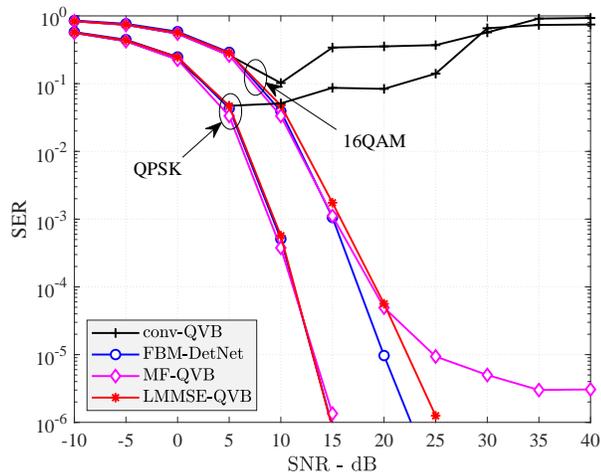}
    \caption{Data detection performance comparison for i.i.d. channels with perfect CSI, $b=3$ bits, $K=16$, $M=32$ and $M=64$ for QPSK and 16QAM signaling, respectively.}
    \label{fig_DD_16K_3bit_iid}
\end{figure}
    
\begin{figure}[t!]
    \centering
    \includegraphics[width=\linewidth]{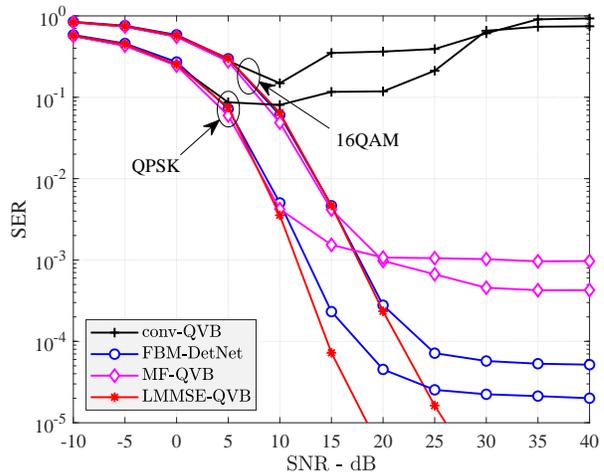}
    \caption{Data detection performance comparison for spatially correlated channels  with perfect CSI,  $b=3$ bits, $K=16$, $M=32$ and $M=64$ for QPSK and 16QAM signaling, respectively.}
    \label{fig_DD_16K_3bit_correlated}
\end{figure}

First, we examine data detection for the case of perfect CSI. Results for i.i.d. and spatially correlated channels are shown in Fig.~\ref{fig_DD_16K_3bit_iid} and Fig.~\ref{fig_DD_16K_3bit_correlated}, respectively. It can be seen that, for both i.i.d. and correlated channels, the conv-QVB method in~\cite{Thoota2021Variational} is outperformed by all other methods and its performance is severely degraded at high SNRs. This is because conv-QVB does not take into account the residual inter-user interference and often encounters the catastrophic cancellation issue at high SNR. For i.i.d. channels, FBM-DetNet, MF-QVB, and LMMSE-QVB all yield the same performance for QPSK signals, while for $16$QAM FBM-DetNet and LMMSE-QVB are similar and both outperform MF-QVB. For spatially correlated channels, LMMSE-QVB provides a significantly lower SER than FBM-DetNet and MF-QVB due to its estimation of the precision matrix $\bs{\Gamma}$ which can better represent the effect of the residual inter-user interference.

\begin{figure}[t!]
    \centering
    \includegraphics[width=\linewidth]{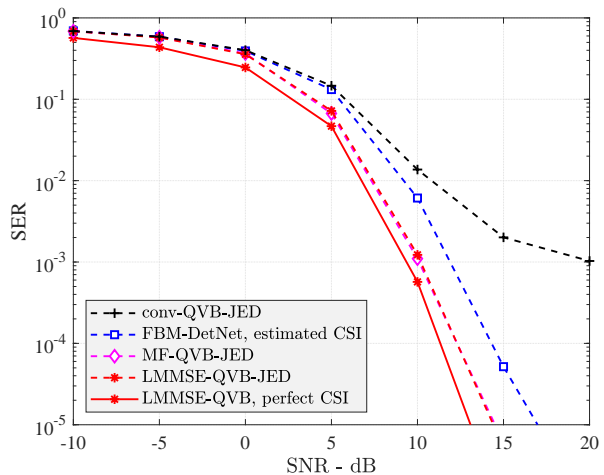}
    \caption{Data detection comparison between the proposed MF-QVB-JED, LMMSE-QVB-JED, and other existing methods for i.i.d. channels with $K = 16$, $M = 32$, $b = 3$ bits, and QPSK signaling.}
    \label{fig_JED_16K_32N_3bit_iid}
\end{figure}

\begin{figure}[t!]
    \centering
    \includegraphics[width=\linewidth]{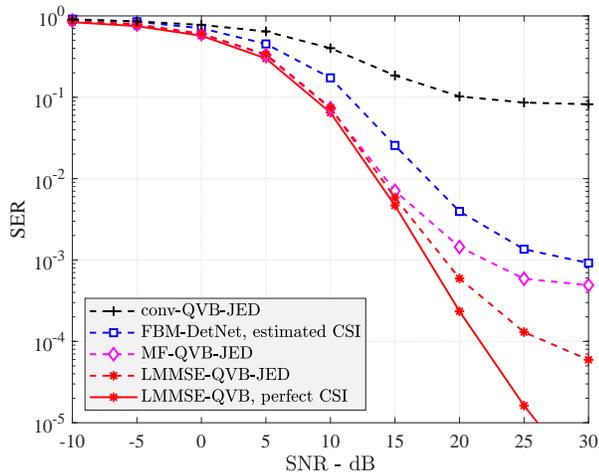}
    \caption{Data detection performance comparison between the proposed MF-QVB-JED, LMMSE-QVB-JED, and other existing methods for spatially correlated channels with $K = 16$, $M = 64$, $b = 3$ bits, and $16$QAM signaling.}
    \label{fig_JED_16K_64N_3bit_correlated}
\end{figure}

Fig.~\ref{fig_JED_16K_32N_3bit_iid} presents results for data detection with estimated CSI and i.i.d. channels. Both MF-QVB-JED and LMMSE-QVB-JED outperform the conventional QVB-JED method as well as the DNN-based detection network FBM-DetNet. Note that FBM-DetNet uses estimated CSI provided by FBM-CENet, a channel estimation network also proposed in~\cite{Ly2022TWC} and designed to estimate the CSI using only the pilot sequence. MF-QVB-JED and LMMSE-QVB-JED both yield the same SER, which is about 2-3dB better than FBM-DetNet at an SER of $10^{-3}$ and $10^{-5}$, respectively. The performance of MF-QVB-JED and LMMSE-QVB-JED is also quite close to that of LMMSE-QVB with perfect CSI.

Results for data detection with estimated CSI and spatially correlated channels are given in Fig.~\ref{fig_JED_16K_64N_3bit_correlated}, where we see that the proposed MF-QVB-JED and LMMSE-QVB-JED methods outperform conv-QVB-JED and FBM-DetNet since the effects of both inter-user interference and spatial channel correlation are taken into account. However, unlike the case of i.i.d. channels where MF-QVB-JED and LMMSE-QVB-JED give the same performance, the LMMSE-QVB-JED method provides a significantly lower SER than MF-QVB-JED at high SNRs for spatially correlated channels. For example, at 30dB, the SER of LMMSE-QVB-JED is about 10 times lower than that of MF-QVB-JED, which is already better than FBM-DetNet. 

    \begin{figure}[t!]
        \centering
        \begin{subfigure}[t]{0.23\textwidth}
            \centering
            \includegraphics[width=\linewidth]{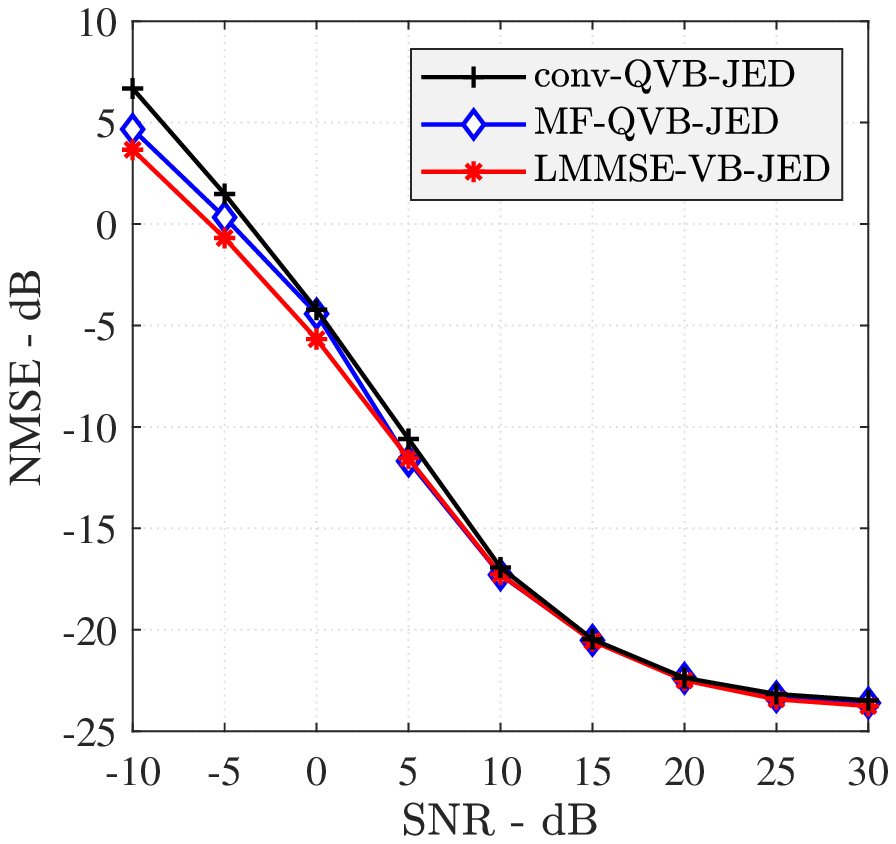}
            \caption{i.i.d. channels, $K = 16$, $M = 32$.}
            \label{fig_CE_16K_32N_iid}
        \end{subfigure}~
        \begin{subfigure}[t]{0.23\textwidth}
            \centering
            \includegraphics[width=\linewidth]{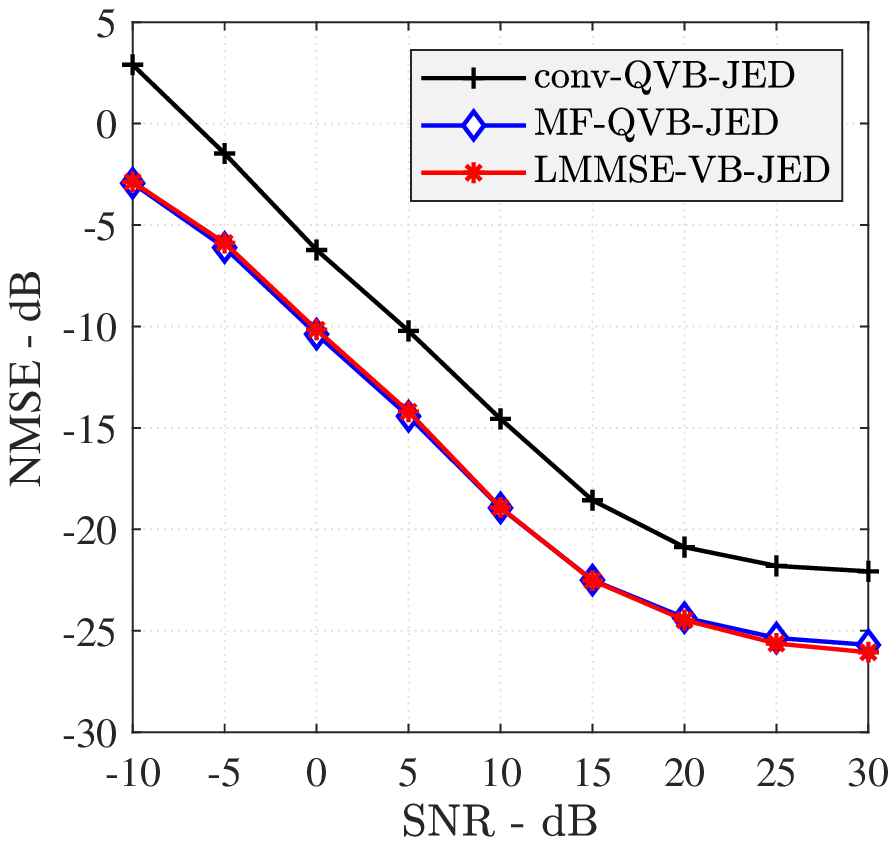}
            \caption{spatially correlated channels, $K = 16$, $M = 64$.}
            \label{fig_CE_16K_64N_correlated}
        \end{subfigure}
        \caption{Channel estimation performance comparison.}
        \label{fig_CE_comparison}
    \end{figure}
    We provide a channel estimation comparison in Fig.~\ref{fig_CE_comparison} where i.i.d. channels are considered in Fig.~\ref{fig_CE_16K_32N_iid} and spatially correlated channels are considered in Fig.~\ref{fig_CE_16K_64N_correlated}. The normalized mean squared error (NMSE) in these figures is defined as $\mr{NMSE} = \mathbb{E}\big[\|\mb{H} - \mb{\hat{H}}\|_F^2/\|\mb{H}\|_F^2\big]$. For i.i.d. channels, all three VB-based methods conv-QVB-JED, MF-QVB-JED, and LMMSE-QVB-JED give similar performance but for spatially correlated channels, the proposed MF-QVB-JED and LMMSE-QVB-JED methods are seen to provide lower NMSEs compared to the conv-QVB-JED method.

    \begin{figure}[t!]
        \centering
        \includegraphics[width=\linewidth]{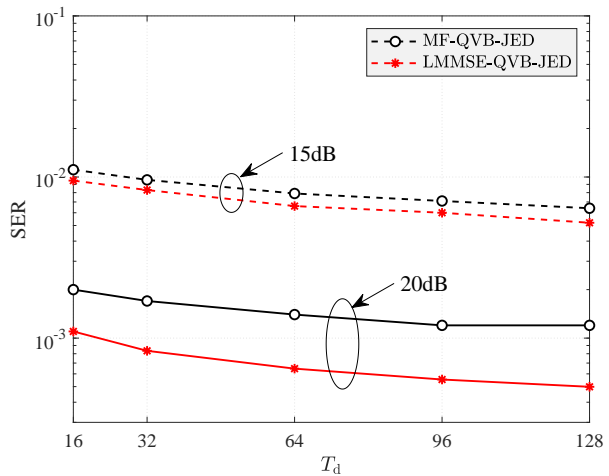}
        \caption{Detection performance of MF-QVB-JED and LMMSE-QVB-JED versus $\Td$ with $K = 16$, $M = 64$, $b = 3$ bits, and 16QAM signaling.}
        \label{fig_SER_vs_Td_16K_64N_3bit_16QAM}
    \end{figure}
    Fig.~\ref{fig_SER_vs_Td_16K_64N_3bit_16QAM} presents the SER performance of the proposed MF-QVB-JED and LMMSE-QVB-JED methods w.r.t. the data transmission length $\Td$. We observe that the SER performance improves with increasing $\Td$ since more received signals are combined to achieve a more accurate channel estimate. Consequently, the data detection phase can result in a lower detection error.

    \begin{figure}[t!]
        \centering
        \includegraphics[width=\linewidth]{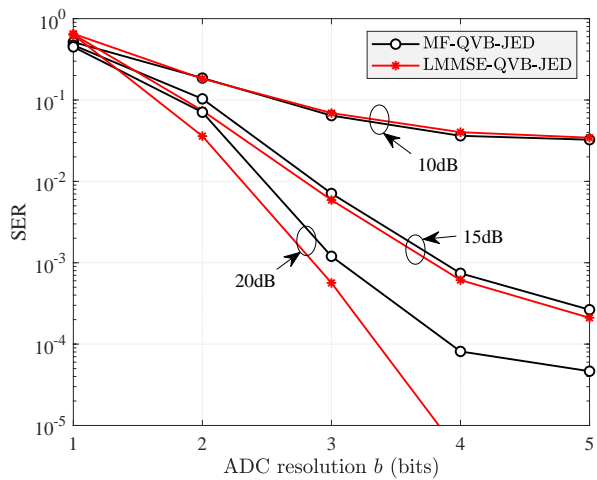}
        \caption{Detection performance of MF-QVB-JED and LMMSE-QVB-JED versus the ADC bit resolution $b$ with $K = 16$, $M = 64$, and 16QAM signaling.}
        \label{fig_SER_vs_bits_16K_64N_16QAM}
    \end{figure}
	In Fig.~\ref{fig_SER_vs_bits_16K_64N_16QAM}, we evaluate the data detection performance of the proposed MF-QVB-JED and LMMSE-QVB-JED methods for different ADC bit resolutions. As expected, increasing the resolution $b$ significantly helps improve the detection performance. It is observed that lower SNRs require a lower bit resolution for the best performance, e.g., $4$-bit ADCs are sufficient to obtain the lowest SER at 10dB. Increasing the ADC bit resolution to values higher than 4 does not result in a lower SER. It is also interesting to note that at high SNRs, LMMSE-QVB-JED can provide much lower SERs compared to MF-QVB-JED as the bit resolution increases.
 
	\section{Conclusion}
	\label{sec:conclusion}
	In this paper, we exploited the VB inference framework to propose different channel estimation and data detection methods for massive MIMO systems with low-resolution ADCs. In particular, we proposed new VB-based algorithms referred to as MF-QVB and LMMSE-QVB for data detection with known CSI, and MF-QVB-JED and LMMSE-QVB-JED for joint channel estimation and data detection. In the proposed QVB framework, we proposed to float the noise variance/covariance matrix as an unknown random variable which also allows the algorithms to take into account the residual inter-user interference. Numerous practical aspects of the QVB framework were studied to improve the implementation stability. It was also shown via a number of simulation studies that the proposed methods provide robust performance and significantly outperform existing methods, particularly when the channels are spatially correlated.

    \appendices
    \section{Proof of Theorem \ref{theorem-1}} \label{append}
    Expanding $\blr{\|\mb{y}-\mb{A}\mb{x}\|^2}$ and taking into account the independence between $\mb{A}$, $\mb{y}$, and $\mb{x}$, we have 
		\begin{align}\label{expand}
			&\blr{(\mb{y}-\mb{A}\mb{x})^H\mb{B}(\mb{y}-\mb{A}\mb{x})} \nonumber \\
			&= \lr{\mb{y}^H\mb{B}\mb{y}} - 2\,\Re\big\{\lr{\mb{y}^H\mb{BAx}}\big\} + \lr{\mb{x}^H\mb{A}^H\mb{B}\mb{A}\mb{x}} \nonumber \\ 
			&= (\lr{\mb{y}}-\lr{\mb{A}}\lr{\mb{x}})^H\mb{B}(\lr{\mb{y}}-\lr{\mb{A}}\lr{\mb{x}})  + \lr{\mb{y}^H\mb{B}\mb{y}} \nonumber \\
			&\quad - \lr{\mb{y}^H}\mb{B}\lr{\mb{y}} +\lr{\mb{x}^H\mb{A}^H\mb{B}\mb{A}\mb{x}} - \lr{\mb{x}^H}\lr{\mb{A}^H}\mb{B}\lr{\mb{A}}\lr{\mb{x}}.
		\end{align}
		Note that $\lr{\mb{x}\mb{x}^H} = \lr{\mb{x}}\lr{\mb{x}}^H + \bs{\Sigma}_{\mb{x}}$ and 
		$\lr{\mb{y}^H\mb{B}\mb{y}} = \tr\{\mb{B}\lr{\mb{y}\mb{y}^H}\} =  \lr{\mb{y}^H}\mb{B}\lr{\mb{y}} +\tr\{\mb{B}\bs{\Sigma_{\mb{y}}}\}$. In addition, we have
	\begin{align}
	\!\!\! \big[\lr{\mb{A}^H\mb{B}\mb{A}}\big]_{ij} &= \lr{\mb{a}_i^H\mb{B}\mb{a}_j} \nonumber \\
		&=
		\left\{ \begin{array}{ll}
			\lr{\mb{a}_i^H}\mb{B}\lr{\mb{a}_i} + \tr\{\mb{B}\mb{\Sigma}_{\mb{a}_i}\}, & \text{if}\; i = j \\
			\lr{\mb{a}_i^H}\mb{B}\lr{\mb{a}_j}, &  \text{otherwise}.\\
		\end{array} \right.
	\end{align}
		It thus follows that $\lr{\mb{A}^H\mb{B}\mb{A}} = \lr{\mb{A}^H}\mb{B}\lr{\mb{A}} + \mb{D}$, and 
		\begin{eqnarray*}
			\lr{\mb{x}^H\mb{A}^H\mb{B}\mb{A}\mb{x}} &=& 	\tr\big\{\lr{\mb{A}^H\mb{B}\mb{A}}\lr{\mb{x}\mb{x}^H}\big\}  \\
			&=& \tr\big\{\lr{\mb{A}^H}\mb{B}\lr{\mb{A}}\lr{\mb{x}}\lr{\mb{x}^H}\big\} + \lr{\mb{x}}^H\mb{D}\lr{\mb{x}}  \\
			&& + \tr\big\{\bs{\Sigma}_{\mb{x}}\mb{D}\big\} +  \tr\big\{\bs{\Sigma}_{\mb{x}}\lr{\mb{A}^H}\mb{B}\lr{\mb{A}}\big\}.
		\end{eqnarray*}
		The statement \eqref{f-ABC} thus follows by removing the duplicated terms in \eqref{expand}. Note that $\lr{\mb{x}}^H\mb{D}\lr{\mb{x}}  + \tr\big\{\bs{\Sigma}_{\mb{x}}\mb{D}\big\}$ can also be written as
		$\lr{\mb{x}}^H\mb{D}\lr{\mb{x}}  + \tr\big\{\bs{\Sigma}_{\mb{x}}\mb{D}\big\} = \sum_{i=1}^n\lr{|x_i|^2}\tr\{\mb{B}\bs{\Sigma_{\mb{a}_i}}\}$.
	
    \section{Computation of $\ms{F}_r(\mu,\gamma,a,b)$ and $\ms{G}_r(\mu,\gamma,a,b)$} \label{append-A}
    For ease of presentation, we denote
     	\begin{align} 
     	\alpha = \sqrt{2\gamma}(a-\mu),\quad 
     	\beta= \sqrt{2\gamma}(b-\mu).
     \end{align}
    For an arbitrary complex random variable $\mc{CN}(\mu,\gamma^{-1})$ whose real and imaginary parts are both truncated on the interval $(a,b)$, the mean $\ms{F} _r(\mu,\gamma,a,b) $ and variance $\ms{G} _r(\mu,\gamma,a,b) $ are computed as
     \begin{align}
     	\ms{F} _r(\mu,\gamma,a,b) &= \mu - \frac{1}{\sqrt{2\lr{\gamma}}} \frac{ \phi(\beta)- \phi(\alpha)}{\Phi(\beta) - \Phi(\alpha)} \\
     	\ms{G} _r(\mu,\gamma,a,b) &=  \frac{1}{2\lr{\gamma}} \Bigg[1 - \frac{\beta_m\phi(\beta) - \alpha\phi(\alpha)}{\Phi(\beta) - \Phi(\alpha)} \nonumber \\ 
     	&\quad \quad\quad\quad\,\;- \left(\frac{\phi(\beta) - \phi(\alpha)}{\Phi(\beta) - \Phi(\alpha)}  \right)^2\Bigg],
     \end{align}
    where the PDF and CDF operators $\phi(\cdot)$ and $\Phi(\cdot)$, as well as the multiplication, division, and square operations are applied individually on the real and imaginary components.  The variance $\ms{G} _r(\mu,\gamma,a,b)$ is computed by adding the variances of the two components.
    
    \section{Computation of $\ms{F}_x(z,\gamma)$ and $\ms{G}_x(z,\gamma)$} \label{append-B}
    Given $z = x + \mc{CN}(0,\gamma^{-1})$, the posterior distribution of $x$ given $z$ is
    \begin{align*}
    p(x|z;\gamma) \propto p(x)\,\mc{CN}(z;x,\gamma^{-1})
    \end{align*}
    For $a \in \mc{S}$, we have
    \begin{align*}
    p(x=a|z;\gamma) = \frac{1}{Z}p_a\,\mr{exp}\big(-\gamma|z-a|^2\big),
    \end{align*}
    where  $Z = \sum_{b\in\mc{S}}p_b\,\mr{exp}\big(-\gamma|z-b|^2\big)$ is a normalization factor. The corresponding posterior mean $\ms{F} _x(z,\gamma)$ and variance $\ms{G} _x(z,\gamma)$ are computed as
    \begin{align*}
    	\ms{F} _x(z,\gamma) &= \sum_{a\in\mc{S}} a\times p(x=a|z,\gamma)  \nonumber \\
    	\ms{G} _x(z,\gamma) & = \sum_{a\in\mc{S}} |a|^2\times p(x=a|z,\gamma) - |\ms{F} _x(z,\gamma)|^2.
    \end{align*}
    
    We note that $\mathbb{E}\big[|x|^2|z;\gamma\big]$ is equal to $|a|^2$ for PSK signaling with transmit energy $|a|^2$, as shown below:
    \begin{align*}
        \mathbb{E}\big[|x|^2|z;\gamma\big] &= \sum_{a\in\mc{S}}|a|^2\frac{1}{Z}p_a\,\mr{exp}\big(-\gamma|z-a|^2\big) \\
        &= |a|^2 \frac{\sum_{a\in\mc{S}}p_a\,\mr{exp}\big(-\gamma|z-a|^2\big)}{Z} \\
        &= |a|^2.
    \end{align*}

	\bibliographystyle{IEEEtran}
	\bibliography{ref.bib}
\end{document}